\documentclass[11pt]{article}
\usepackage[margin=1in]{geometry}
\usepackage[T1]{fontenc}
\usepackage[utf8]{inputenc}
\usepackage{lmodern}
\usepackage{microtype}
\usepackage{amsmath,amssymb,bm}
\usepackage{graphicx}
\usepackage{booktabs}
\usepackage{tabularx}
\usepackage{siunitx}
\usepackage{physics}
\usepackage{hyperref}
\usepackage[numbers,sort&compress]{natbib}
\usepackage{setspace}
\usepackage{authblk}
\usepackage{xcolor}
\usepackage{float}
\usepackage{placeins}

\hypersetup{
  colorlinks=true,
  linkcolor=blue,
  citecolor=blue,
  urlcolor=blue
}

\title{Detector-Grade Germanium as a Low-Disorder Host for Indium-Acceptor Spin Qubits: A Five-Qubit Materials-to-Architecture Design Study}

\author[1,*]{D.-M. Mei}
\author[1]{K.-M. Dong}
\author[1]{S. A. Panamaldeniya}
\author[1]{N. Budhathoki}
\author[1]{S. Chhetri}
\author[1]{A. Prem}
\author[1]{S. Bhattarai}

\affil[1]{Department of Physics, University of South Dakota, Vermillion, South Dakota 57069, USA}
\affil[*]{Corresponding author: \href{mailto:dongming.mei@usd.edu}{dongming.mei@usd.edu}}

\date{\today}

\begin{document}

\maketitle

\begin{abstract}
Acceptor-bound hole spins in germanium (Ge) provide a promising but comparatively underexplored route to semiconductor quantum information processing. Here we present a theory-guided design study, not an experimental demonstration, of a detector-grade Ge acceptor-spin platform based on intentionally incorporated indium (In) acceptors in ultra-high-purity Ge. The proposed materials strategy combines a residual impurity background near $10^{10}\,\mathrm{cm^{-3}}$ with a target In density of approximately $2\times10^{14}\,\mathrm{cm^{-3}}$, giving a characteristic three-dimensional acceptor spacing of about \SI{170}{\nano\meter}. A \SI{1}{\micro\meter}-long active channel with a suitable transverse mode volume can therefore contain about five acceptors on average, enabling a statistically selected post-fabrication register rather than a deterministically placed chain. We analyze the physical basis, device architecture, strain and disorder limits, coupling hierarchy, multiscale modeling workflow, fabrication pathway, and scaling prospects of this platform. The central conclusion is that detector-grade Ge can, in principle, suppress uncontrolled bulk electrostatic and strain disorder to levels compatible with acceptor-hole qubits, while the spin--orbit-active valence-band manifold retains all-electrical control and dipolar or phonon-mediated coupling channels. Direct exchange is treated as a close-pair or gate-enhanced channel, not as the generic mean-spacing interaction. Phononic crystal cavity engineering is therefore best viewed as a second-stage enhancement that can suppress unwanted acoustic modes and enable selected cavity-mediated interactions after baseline control, readout, and nearest-neighbor coupling are validated. The principal remaining challenges are statistical acceptor placement, interface-induced disorder, charge-noise control, readout integration, and experimental validation. These results identify detector-grade Ge In-acceptor qubits as a credible intermediate architecture between donor-based impurity qubits and fully gate-defined Ge hole-spin hardware.
\end{abstract}

\section{Introduction}

The central challenge in quantum computing is no longer the demonstration of an isolated high-fidelity qubit, but the development of a hardware architecture that can scale toward useful and eventually fault-tolerant operation. In practice, scalability requires simultaneously reducing physical error rates, maintaining device uniformity and connectivity, and preventing the classical control stack from overwhelming the cryostat through wiring density, power dissipation, and thermal load \cite{Scappucci2021,Hendrickx2021Nature}. For semiconductor spin qubits, the key question is therefore not only how to preserve coherence, but how to do so in a manufacturable, densely integrated, and electrically controllable device platform.

Among semiconductor hosts, high-purity germanium (Ge) has emerged as one of the most attractive materials for spin-based quantum information processing \cite{Scappucci2021}. Ge combines small carrier effective masses, high mobilities, strong but engineerable spin--orbit coupling, compatibility with isotopic purification, and mature semiconductor processing. In particular, the spin-$3/2$ valence-band manifold supports strong electric-dipole spin resonance (EDSR), electrically tunable anisotropic $g$ tensors, and efficient all-electrical control \cite{Hendrickx2020Fast,Watzinger2018}. These advantages have already made gate-defined Ge hole-spin qubits one of the leading semiconductor platforms for near-term quantum processors, with demonstrations ranging from ultrafast coherent control and two-qubit logic to four-qubit operation and coherent spin shuttling \cite{Hendrickx2020Fast,Hendrickx2020,Hendrickx2021Nature,Wang2022Hole,vanRiggelen2024,Hendrickx2024Sweet}.

Gate-defined dots, however, are not the only route to Ge-based quantum computing. Acceptor-bound hole spins provide a distinct modality in which confinement is defined primarily by chemistry rather than by a complex electrostatic gate pattern. This distinction matters because lithographically defined dots, while highly flexible, often require substantial gate complexity and are sensitive to interface disorder, alloy fluctuations, and local strain inhomogeneity. Acceptor qubits instead offer chemically defined localization, potentially improved site reproducibility, and natural compatibility with compact one-dimensional or two-dimensional arrays.

Acceptor qubits are particularly attractive in Ge because the same valence-band physics that benefits gate-defined hole qubits also applies to impurity-bound holes. A substitutional acceptor binds a valence hole whose low-energy manifold inherits spin-$3/2$ heavy-hole/light-hole (HH/LH) structure, strong spin--orbit interaction, quadrupolar couplings, and pronounced sensitivity to electric fields and strain \cite{AbadilloUriel2016,AbadilloUriel2017}. At cryogenic temperatures, where thermal ionization of acceptor-bound holes is strongly suppressed, the acceptor-bound hole can form an electrically active dipolar state relative to its ionized acceptor core, providing a physical basis for electric-field control and coupling to phonons or cavity fields. Operation near \SI{4}{K} is treated here as a design target rather than as an experimentally demonstrated regime for coherent GHz In-acceptor spin-qubit operation. The main challenge is therefore not only whether a qubit degree of freedom exists, but whether the host material and device environment can be made sufficiently clean and uniform to support coherent control without the acceptor manifold being dominated by uncontrolled strain, charge disorder, or interface fields.

This is where detector-grade Ge becomes especially important. Through zone refining and Czochralski growth, residual impurity concentrations can be reduced to the $10^{10}\,\mathrm{cm^{-3}}$ scale, corresponding to practical detector-grade material with extremely low electrostatic and strain disorder \cite{Hansen1982,Wang2012HPGe,Wang2014Dislocation}. Such materials quality is substantially beyond conventional electronic-grade Ge and may be essential for realizing stable acceptor-spin qubits.

In this paper, we propose a detector-grade Ge acceptor-spin architecture based on intentionally introduced indium (In) acceptors. The central idea is to first suppress the residual impurity background to approximately $10^{10}\,\mathrm{cm^{-3}}$, then introduce a controlled In concentration near $2\times10^{14}\,\mathrm{cm^{-3}}$ so that the intentional qubit-defining acceptors dominate the local environment while random background disorder remains negligible. At this density, the three-dimensional mean acceptor spacing is naturally on the order of 150--200 nm. A \SI{1}{\micro\meter}-long active channel with an appropriate transverse mode volume can therefore contain, on average, approximately five acceptors, suggesting a compact statistically selected five-site register suitable for proof-of-principle devices and short quantum buses.

The specific novelty of the present work is not the generic idea of acceptor qubits in Ge, but the integrated materials-to-architecture proposal: detector-grade Ge as the low-disorder host, controlled In incorporation as the qubit-defining strategy, a statistically selected one-dimensional five-site register as the minimal device unit, and an engineered phononic crystal cavity (PnC) as a second-stage enhancement for suppressing unwanted environmental phonons and enabling selected cavity-mediated coupling. The scope of the paper is therefore that of a design-and-architecture study, supported by order-of-magnitude analysis and multiscale modeling logic, rather than an experimental device demonstration.

The goal of this paper is to present the physical basis, architecture, modeling framework, fabrication pathway, and scaling logic of this platform. We first establish the design principles of the five-qubit chain and the materials physics of In acceptor qubits in detector-grade Ge. We then analyze residual strain and disorder, spin--phonon coupling and coupling mechanisms, a multiscale modeling workflow, and fabrication routes compatible with high-purity Ge growth and semiconductor device processing. Finally, we assess scalability toward longer one-dimensional buses and modular arrays, and compare the platform with donor-spin and gate-defined Ge qubit approaches.

To keep the scope explicit, the quantitative statements in this paper fall into three categories. First, \emph{literature-grounded inputs} include the established Ge hole-spin materials physics, detector-grade impurity scale, and measured In incorporation behavior. Second, \emph{design-level estimates derived here} include the characteristic acceptor spacing, the five-qubit-per-\SI{1}{\micro\meter} design regime, and the residual strain/disorder scaling arguments. Third, \emph{nominal modeling outputs} include the gate-response, dipole-coupling, and PnC-enhanced performance ranges used to assess architectural plausibility. This distinction is important: the present manuscript aims to establish a credible design window and a staged development pathway, not to claim full device-specific validation. These claim categories are summarized in Table~\ref{tab:claim_status}.

\begin{table}[H]
\centering
\caption{Status of the principal quantitative claims used in this design-and-architecture study.}
\label{tab:claim_status}
\begin{tabular}{p{3.3cm}p{4.7cm}p{5.2cm}}
\toprule
\textbf{Category} & \textbf{Representative examples} & \textbf{Role in the paper} \\
\midrule
Literature-grounded inputs & Ge hole-spin control physics, detector-grade impurity scale, measured In segregation behavior & Physical basis and externally established constraints \\
Design-level estimates in this work & $d_{\mathrm{avg}}\!\approx\!\SI{170}{nm}$, statistically selected $\sim 5$-site active volume, impurity-strain scaling, disorder-floor estimates & Architecture definition and feasibility window \\
Nominal modeling outputs & Gate crosstalk hierarchy, dipolar/exchange/phonon coupling ranges, engineered PnC suppression targets & Internal consistency checks for the proposed device concept \\
\bottomrule
\end{tabular}
\end{table}

\section{Five-Qubit Device Architecture in a One-Dimensional \texorpdfstring{$\SI{1}{\micro\meter}$}{1 um} Platform}

The proposed architecture is based on a simple design principle: use detector-grade Ge as an ultra-low-disorder host crystal and define the qubits through intentionally introduced indium (In) acceptors rather than through fully lithographic electrostatic quantum dots. In this approach, the primary burden of qubit definition is shifted away from complex multi-gate confinement and toward materials control during crystal growth and doping, where stronger mesoscopic uniformity can in principle be achieved. At the same time, the platform is designed not only as an electrostatically tunable acceptor-spin chain, but also as a phonon-engineered device in which the local acoustic environment can be structured to suppress unwanted environmental phonons and, when desired, support selected cavity-assisted spin--phonon interactions \cite{Smelyanskiy2014GePhononic,Mei2025QST}.

We consider detector-grade Ge first purified to a residual background impurity concentration near \(n_{\mathrm{bg}}\sim10^{10}~\mathrm{cm^{-3}}\), followed by controlled In incorporation to a target acceptor concentration of \(n_{\mathrm{In}}\sim2\times10^{14}~\mathrm{cm^{-3}}\). These quantities are introduced here only as design inputs and are formally defined later in Eqs.~\eqref{eq:nbg} and \eqref{eq:nin}. Because the intentional acceptor density exceeds the random residual background by roughly four orders of magnitude, the local electrostatic and strain environment is expected to be set primarily by the designed acceptor ensemble rather than by uncontrolled background impurities.

At this target concentration, the mean three-dimensional acceptor spacing is approximately \(d_{\mathrm{avg}}\sim n_{\mathrm{In}}^{-1/3}\approx\SI{170}{\nano\meter}\), as derived in Eq.~\eqref{eq:davg}.

This spacing is large enough to preserve atom-like localization of individual acceptor-bound holes and is most naturally compatible with electrically tunable dipolar coupling or phonon-assisted coupling. It should not be interpreted as implying strong direct wavefunction-overlap or exchange coupling at the mean spacing. Because a plausible acceptor Bohr radius in Ge is only \(a_B^*\sim 2\)--\(5\,\mathrm{nm}\), direct tunnel or exchange coupling would require much smaller selected separations, typically on the order of several \(a_B^*\), roughly \(10\)--\(30\,\mathrm{nm}\), or substantial gate-induced hybridization. The estimate should be interpreted statistically rather than deterministically: \(d\sim n_{\mathrm{In}}^{-1/3}\) is a three-dimensional mean-spacing estimate and does not imply that five acceptors automatically lie on a perfectly straight one-dimensional line. A usable five-site chain requires an effective active channel volume selected by the gate geometry, electrostatic confinement, and charge-sensing window. For \(n_{\mathrm{In}}\sim 2\times10^{14}\,\mathrm{cm^{-3}}\), a \(\SI{1}{\micro\meter}\)-long active channel with a transverse mode area of order \((150\text{--}170\,\mathrm{nm})^2\) contains approximately five acceptors on average. As quantified by the Poisson occupancy estimate in Section~\ref{sec:design_modeling_workflow}, a nominal mean occupancy of five gives \(P(N=5)\simeq0.18\), \(P(4\leq N\leq6)\simeq0.50\), and \(P(3\leq N\leq7)\simeq0.74\) before applying device-usability constraints. The first-generation device should therefore be viewed as a statistical-yield architecture: random acceptor configurations are produced by controlled chemical incorporation, and usable chains are identified, mapped, and tuned electrically after fabrication. Crystal-growth parameters can control the mean In concentration and its axial/radial profile, but the microscopic acceptor positions remain statistical; the proposed register therefore relies on post-fabrication mapping and selection rather than deterministic placement of individual In atoms.

In practice, the device workflow would proceed by screening nominal five-site regions using low-temperature charge sensing, Stark-shift spectroscopy, and EDSR response. Local plunger gates would then select a subset of acceptors with suitable spacing, tunability, and coupling hierarchy, while barrier gates would tune the electrostatic environment between neighboring sites. Electrical mapping through RF charge sensing and gate spectroscopy should be interpreted primarily as an effective electrostatic map of the active acceptors: relative gate lever arms can help infer lateral position, whereas depth below the gate generally requires electrostatic modeling or calibration from companion characterization structures. This approach avoids the immediate requirement of deterministic single-ion implantation, but it also means that early devices should be evaluated by a design-yield metric rather than by assuming a perfectly deterministic register. Deterministic implantation or atomically precise placement could later be explored as a higher-control prototype route, but it is not required for the baseline statistical-chain demonstration proposed here.

A central extension introduced here is to embed this acceptor chain within, or adjacent to, a patterned phononic crystal (PnC) environment. The purpose of the PnC is twofold. First, by opening an acoustic bandgap over the frequency range relevant to spontaneous qubit relaxation, it can strongly suppress emission into unwanted environmental phonon modes. Second, by introducing a defect cavity or guided phononic channel, it can preserve selected localized or propagating acoustic modes for controlled qubit--qubit coupling or spin--phonon transduction \cite{Smelyanskiy2014GePhononic,Mei2025QST}. The phonon-limited relaxation rate may be written schematically as \(\Gamma_1^{\mathrm{ph}}\propto |g_{\mathrm{sp}}|^2\rho_{\mathrm{ph}}(\omega_q)\), with the formal expression given later in Eq.~\eqref{eq:Gamma1_DOS}; reducing the phononic density of states at the qubit frequency directly suppresses unwanted one-phonon decay channels.
\begin{figure}[htp!]
    \centering
    \includegraphics[width=\textwidth]{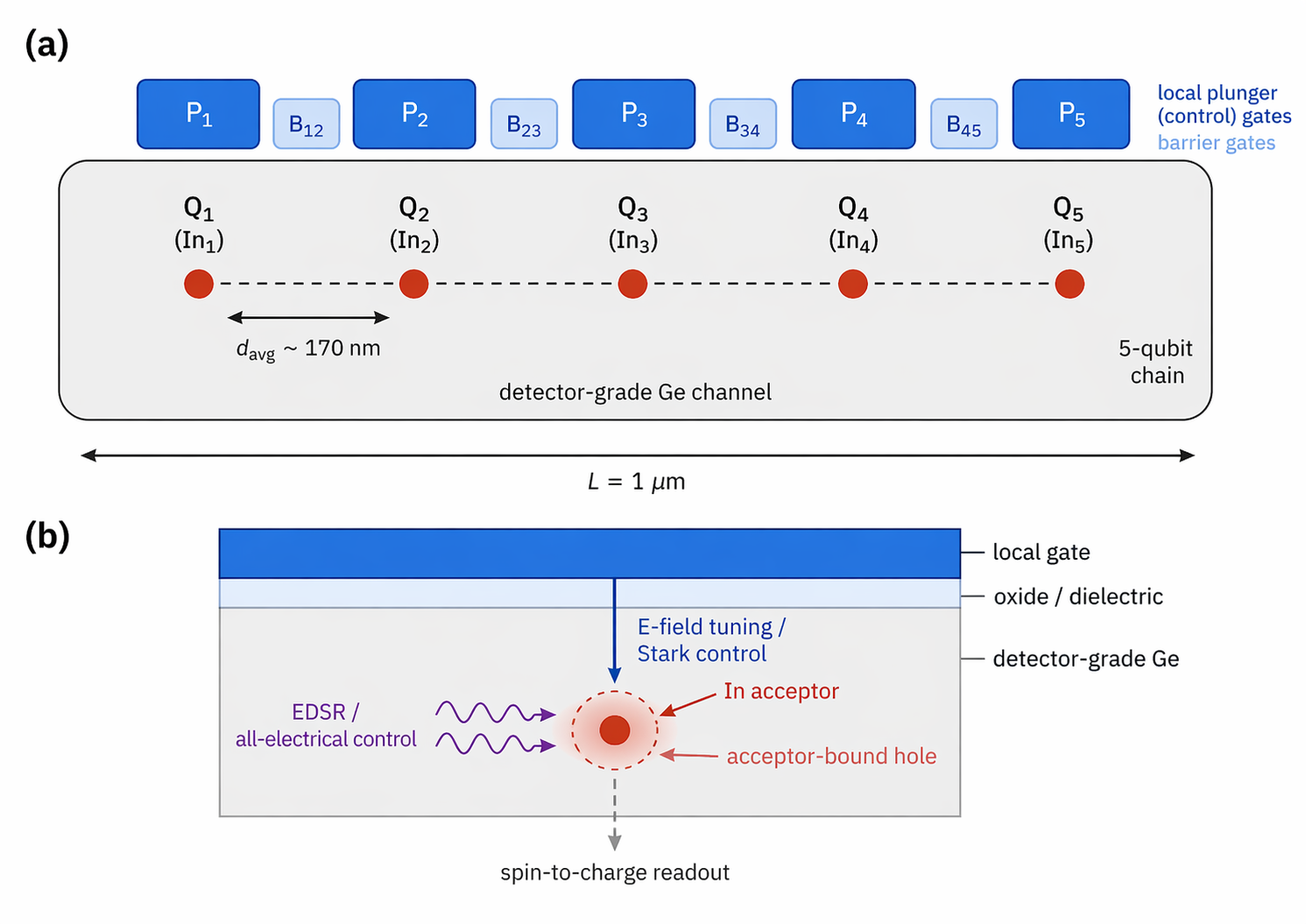}
    \caption{Conceptual design of a five-qubit indium-acceptor platform in detector-grade Ge. \textbf{(a)} Top-view architecture of a one-dimensional qubit chain formed by five mapped In acceptor sites, labeled $Q_1$--$Q_5$, embedded in a detector-grade Ge channel. The labels $Q_1$--$Q_5$ represent one successfully mapped post-fabrication realization of the statistical acceptor distribution, not predetermined implantation sites. A shallow surface gate stack provides local plunger/control gates ($P_1$--$P_5$) and barrier gates ($B_{12}$, $B_{23}$, $B_{34}$, $B_{45}$) that are biased after mapping to tune the local electrostatic environment and inter-qubit coupling. In a practical device, additional outer barrier or reservoir-coupling gates may be included to tune coupling between the end acceptors and the source/drain reservoirs; these gates are omitted here for schematic clarity. In an improved implementation, the Ge channel is surrounded by a patterned phononic crystal region that suppresses unwanted environmental acoustic modes while allowing selected cavity or guided modes to be engineered for controlled coupling. At the target acceptor concentration $n_{\mathrm{In}}\sim 2\times10^{14}~\mathrm{cm^{-3}}$, the three-dimensional mean acceptor spacing is $d_{\mathrm{avg}}\sim \SI{170}{\nano\meter}$; a \SI{1}{\micro\meter}-long active channel with a suitable transverse mode volume can therefore host, on average, a statistically selected five-site register. \textbf{(b)} Cross-sectional conceptual view of an individual qubit beneath a local gate. The gate electric field provides Stark tuning and enables all-electrical control through electric-dipole spin resonance (EDSR), while the surrounding phononic environment is engineered to filter unwanted phonons and, in a cavity-enabled design, to localize selected acoustic modes near the qubit. The physical qubit is hosted by an acceptor-bound hole localized around a substitutional In acceptor in detector-grade Ge, with spin-to-charge readout implemented through a nearby RF-QPC, RF-SET, or gate-based dispersive charge sensor after a spin-dependent mapping pulse.}
    \label{fig:five_qubit_architecture}
\end{figure}

Figure~\ref{fig:five_qubit_architecture} shows the conceptual layout of this one-dimensional five-qubit platform. The active region consists of a detector-grade Ge channel in which five usable In acceptor sites are statistically selected within a \SI{1}{\micro\meter}-scale gate-defined sensing window. The schematic labels $Q_1$--$Q_5$ should therefore be understood as one mapped and usable post-fabrication realization of the random acceptor distribution rather than as predefined impurity positions. A shallow surface gate stack, represented by local plunger/control gates, provides electric-field tunability over the underlying acceptor sites. In the simplest implementation, these gates tune the local Stark shift, modify the effective dipole moment of each acceptor-bound hole, and bring selected qubits into or out of resonance for electric-dipole spin resonance (EDSR)-based control.

In the phonon-engineered version of the platform, the one-dimensional chain is embedded within a PnC membrane or surrounded laterally by a patterned acoustic shield. The simplest operating mode uses the PnC primarily as a coherence-preserving environment: qubit frequencies are placed inside an acoustic bandgap so that the dominant unwanted one-phonon relaxation channels are suppressed. A more advanced operating mode introduces a designed defect cavity or phononic waveguide so that selected acoustic modes remain available as resources rather than uncontrolled loss channels.

The physical qubit is encoded in the lowest Kramers doublet of the acceptor-bound hole. The nearby excited doublet provides the orbital admixture needed for electrically driven control, so that the system retains the key advantage of Ge hole-based quantum devices---strong spin--orbit-assisted all-electrical manipulation---while avoiding the full gate complexity of a conventional quantum-dot processor. The result is a hybrid architecture that combines atom-defined localization with gate-tunable qubit frequencies and couplings, while using phononic band engineering to suppress unwanted environmental phonons.

This five-qubit geometry is therefore best viewed as the minimal experimentally meaningful building block of the proposed platform. It is large enough to study individual control, frequency allocation, nearest-neighbor coupling, crosstalk, and short-range quantum transport, while also providing a natural testbed for phonon filtering, cavity protection, and selected spin--phonon coupling in detector-grade Ge. More importantly, it establishes the basic module from which longer one-dimensional quantum buses and more complex modular Ge-based acceptor-spin architectures can be developed.

\section{Physics Basis of In Acceptor Qubits in Detector-Grade Ge}
\label{sec:physics_in_acceptor}

Detector-grade germanium offers an unusually favorable host for acceptor-spin qubits because it combines three key ingredients: a chemically and electrically ultra-clean crystal, a valence-band manifold with strong spin--orbit coupling, and a host nuclear environment that is already comparatively quiet even before isotopic enrichment \cite{Scappucci2021}. In natural Ge, only $^{73}$Ge carries nuclear spin, while valence-band holes have predominantly $p$-like orbital character and therefore experience a strongly suppressed Fermi-contact hyperfine interaction relative to conduction electrons \cite{Scappucci2021}. These advantages are already reflected in the rapid progress of gate-defined Ge hole-spin qubits, including the demonstration of single-hole operation, a four-qubit processor, ultrafast coherent control, and sweet-spot operation with strongly improved coherence \cite{Hendrickx2020,Hendrickx2021Nature,Wang2022Hole,Hendrickx2024Sweet}. The present proposal seeks to transfer those materials advantages from gate-defined dots to an atom-like impurity platform based on indium (In) acceptors in detector-grade Ge.

A second enabling feature is the exceptionally high chemical purity achievable in detector-grade Ge. High-purity Ge crystals intended for radiation detectors are required to have net electrically active impurity concentrations on the order of $10^{10}~\mathrm{cm^{-3}}$ or below \cite{Hansen1982}. Modern zone-refining and Czochralski-growth programs have demonstrated high-purity Ge crystals in this regime together with dislocation densities compatible with detector operation and advanced device fabrication \cite{Wang2012HPGe,Wang2014Dislocation}. For acceptor qubits, this matters directly: acceptor-bound holes are highly sensitive to local strain, symmetry breaking, and electrostatic disorder, so detector-grade Ge suppresses the uncontrolled background that would otherwise broaden the qubit spectrum and destabilize the qubit-to-qubit frequency map. Against this materials backdrop, the central physical question becomes how the acceptor manifold, its electric-dipole character, and its phonon environment can be engineered into a coherent qubit platform.

\subsection{Acceptor-bound-hole manifold and qubit encoding}

A substitutional group-III impurity in Ge binds a valence hole whose low-energy structure derives from the $\Gamma$-point spin-$3/2$ heavy-hole/light-hole manifold. In bulk, the acceptor ground state is approximately fourfold degenerate, but in any realistic device this degeneracy is lifted by interface confinement, dielectric mismatch, strain, and gate-induced electric fields \cite{AbadilloUriel2016,AbadilloUriel2017}. A compact starting Hamiltonian may therefore be written as
\begin{equation}
H_{\mathrm{acc}}
=
H_{\mathrm{KL}}
+
V_{C}(\mathbf{r})
+
V_{\mathrm{cc}}
+
H_{\epsilon}
+
e\,\mathbf{E}\!\cdot\!\mathbf{r}
+
H_{B},
\label{eq:Hacc_full}
\end{equation}
where $H_{\mathrm{KL}}$ is the Kohn--Luttinger valence-band Hamiltonian, $V_C$ is the screened Coulomb potential of the acceptor core, $V_{\mathrm{cc}}$ is the central-cell correction, $H_{\epsilon}$ describes strain-induced heavy-hole/light-hole mixing, $e\,\mathbf{E}\!\cdot\!\mathbf{r}$ is the electric-field coupling, and $H_B$ is the Zeeman interaction.

\begin{figure}[H]
    \centering
    \includegraphics[width=0.98\textwidth]{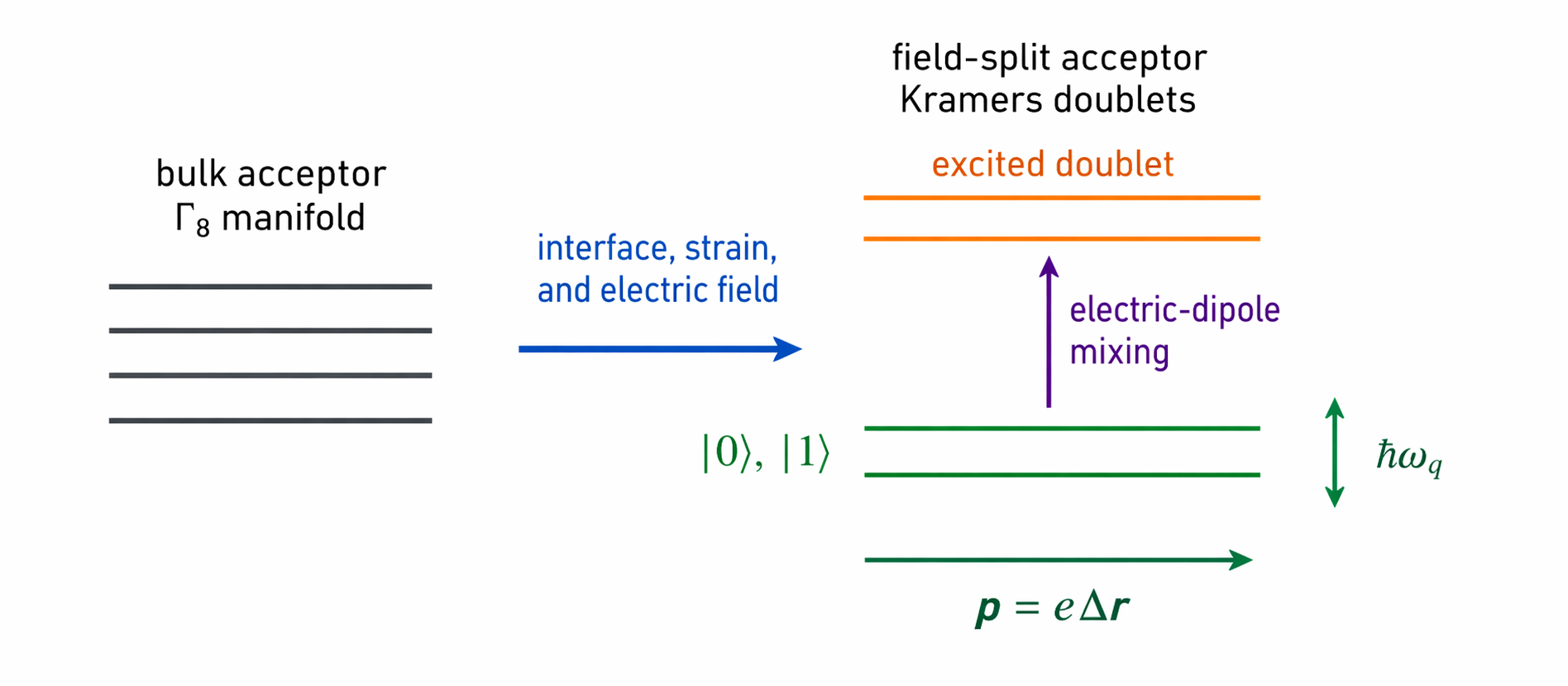}
    \caption{Schematic physics of an In acceptor qubit in Ge. In bulk, the acceptor-bound hole inherits the fourfold $\Gamma_8$ valence-band manifold. Near an interface or under an applied electric field, this manifold splits into two Kramers doublets. The lowest doublet forms the qubit basis, while the excited doublet enables electric-dipole-mediated control through spin--orbit-induced mixing.}
    \label{fig:acceptor_doublet}
\end{figure}

Near an interface, the fourfold acceptor manifold splits into two Kramers doublets and inversion-symmetry breaking induces parity mixing \cite{AbadilloUriel2016}. The lowest doublet then defines the computational basis,
\begin{equation}
\{|0\rangle,\ |1\rangle\},
\end{equation}
while the excited doublet provides the nearby orbital structure that enables electrically driven spin manipulation. After projection into the lowest Kramers doublet, the qubit Hamiltonian takes the generic form
\begin{equation}
H_{\mathrm{q}}
=
\frac{\mu_B}{2}\,\mathbf{B}\!\cdot\!\mathbf{g}(\mathbf{E},\epsilon)\!\cdot\!\bm{\sigma}
+
\frac{\hbar \Omega_R}{2}\cos(\omega t)\,\sigma_x
+
\frac{\delta\omega_z(t)}{2}\,\sigma_z,
\label{eq:Hq_eff}
\end{equation}
where $\mathbf{g}(\mathbf{E},\epsilon)$ is the effective, electrically tunable $g$-tensor, $\Omega_R$ is the electrically driven Rabi frequency, and $\delta\omega_z(t)$ represents slow fluctuations arising from charge noise, strain noise, or residual hyperfine fields. Equation~\eqref{eq:Hq_eff} makes clear why acceptor holes are attractive qubits in Ge: the same spin--orbit-active valence-band physics that drives Ge hole-dot qubits also enables all-electrical control of impurity-bound holes. Figure~\ref{fig:acceptor_doublet} shows a schematic of the corresponding In-acceptor qubit physics in Ge.

\subsection{Electric dipole character and all-electrical control}

The qubit Hamiltonian above naturally leads to the next key ingredient of the platform: the acceptor-bound hole is not only spin--orbit-active, but also electrically polarizable. At cryogenic temperature, the acceptor-bound hole remains localized around the ionized acceptor core and acquires an electric dipole moment when interface asymmetry or a local gate field displaces the center of the hole wavefunction relative to the core. A useful definition is
\begin{equation}
\mathbf{p}
=
e\,\Delta \mathbf{r}
=
e\left(
\langle \psi_h|\mathbf{r}|\psi_h\rangle
-
\mathbf{R}_{\mathrm{In}}
\right),
\label{eq:dipole}
\end{equation}
where $\mathbf{R}_{\mathrm{In}}$ is the position of the substitutional In acceptor and $|\psi_h\rangle$ is the bound-hole state. This dipole moment is the microscopic origin of two important features. First, it permits electric-dipole spin resonance through the electric modulation of the spin--orbit-active acceptor manifold. Second, it allows direct dipole--dipole or phonon-mediated interactions between nearby acceptors.

For two acceptors separated by $\mathbf{r}_{ij}=r_{ij}\hat{\mathbf{r}}_{ij}$, the leading electrostatic coupling scale is
\begin{equation}
U_{ij}^{\mathrm{dd}}
=
\frac{1}{4\pi \varepsilon_{\mathrm{Ge}}}
\,
\frac{
\mathbf{p}_i\!\cdot\!\mathbf{p}_j
-
3(\mathbf{p}_i\!\cdot\!\hat{\mathbf{r}}_{ij})
(\mathbf{p}_j\!\cdot\!\hat{\mathbf{r}}_{ij})
}{
r_{ij}^{3}
},
\label{eq:dipole_dipole}
\end{equation}
which is strongly tunable through gate-controlled dipole moments and qubit spacing. In practice, the effective qubit--qubit interaction may be electrostatic, exchange-assisted, or phonon-assisted depending on the overlap of the bound-hole wavefunctions and the local phononic environment. This naturally motivates the next step in the design logic: if acceptor holes are intrinsically strain-active as well as electric-dipole-active, then the acoustic environment itself becomes an important architectural parameter rather than merely an external source of decoherence.

\subsection{Phononic crystal cavity engineering for coherence protection and controlled coupling}

For Ge acceptor qubits, the same spin--orbit-active valence-band physics that enables fast electrical control also makes the qubit sensitive to acoustic strain. This creates a central trade-off: coupling to phonons can enable useful strain-assisted control and spin--phonon interfaces, but it can also open unwanted relaxation and dephasing channels. A natural way to manage this trade-off is to embed the acceptor array in a phononic crystal (PnC) environment, ideally with a defect cavity or waveguide engineered into the surrounding Ge structure \cite{Smelyanskiy2014GePhononic,Mei2025QST}.

At the qubit level, acoustic strain enters through the valence-band strain Hamiltonian and modifies the acceptor doublet through $g$-tensor modulation, orbital mixing, and HH/LH admixture. The formal spin--phonon Hamiltonian, density-of-states dependence of the relaxation rate, and cavity-mediated coupling model are developed in Section~\ref{sec:spin_phonon}. Here we emphasize only the architectural role of the PnC: it reshapes the local acoustic spectrum seen by the qubit rather than defining the qubit itself.

This distinction is important for the staged design strategy. If the qubit transition is placed inside a phononic bandgap, unwanted one-phonon relaxation into the acoustic continuum can be suppressed. Conversely, a deliberately introduced defect mode or guided mode can preserve a selected acoustic channel for controlled spin--phonon coupling. The PnC therefore has two complementary functions: it filters the uncontrolled acoustic bath and, in a later-generation device, supplies selected modes for nonlocal coupling or transduction.

For that reason, the PnC should be treated as a second-stage enhancement rather than a prerequisite for the first five-qubit demonstration. The baseline chain can first test localization, Stark tunability, addressability, spin-to-charge readout, and nearest-neighbor coupling. Once those basic functions are validated, phononic engineering can be added to improve the relaxation-limited coherence ceiling and enable selected cavity-mediated interactions.

\subsection{Why indium is a practical acceptor species in Ge}

With the qubit physics and its electric- and strain-response established, the next question is whether there exists a practical impurity species that can be incorporated in a controlled way into detector-grade Ge. Indium is attractive as the intentional acceptor species because its incorporation in high-purity Ge has already been characterized experimentally in Czochralski-grown crystals \cite{Wang2018In}. In particular, indium-doped Ge crystals grown from highly pure starting material exhibit a measured effective segregation coefficient
\begin{equation}
k_{\mathrm{eff}}(\mathrm{In\ in\ Ge})
\simeq
9\times 10^{-4},
\label{eq:keff_in}
\end{equation}
over a wide concentration range \cite{Wang2018In}. This very small $k_{\mathrm{eff}}$ does \emph{not} by itself imply a perfectly flat concentration profile over an entire boule; on the contrary, careful growth design, wafer selection, and axial-position control are required to obtain a wafer with the desired local In density. However, it does mean that the incorporation behavior is predictable and experimentally calibratable, which is valuable for a qubit platform built from intentionally introduced acceptors rather than random background dopants.

The choice of In should therefore be understood as a materials-and-architecture choice, not simply as a choice of the shallowest acceptor in Ge. The common group-III acceptors B, Al, Ga, and In all produce shallow acceptor levels in Ge, with ionization energies clustered near \(10\)--\(11\,\mathrm{meV}\). In this sense, In is not unique because it is shallow; rather, it is attractive because its incorporation behavior in high-purity Ge is experimentally measurable, highly segregating, and therefore compatible with a wafer-selection strategy in which the local acceptor concentration is calibrated after growth. Table~\ref{tab:groupIII_acceptors_ge} summarizes the relevant comparison.

At the proposed operating temperature near \(4\,\mathrm{K}\), the thermal energy is \(k_{\mathrm{B}}T\approx 0.34\,\mathrm{meV}\), which is much smaller than the In acceptor ionization energy of approximately \(11.2\,\mathrm{meV}\). This comparison primarily supports acceptor freeze-out: the acceptor-bound hole is expected to remain localized unless it is deliberately ionized or hybridized by gate fields, optical excitation, or coupling to a reservoir. It should not be read as evidence by itself for coherent GHz qubit operation, which also requires experimental validation of relaxation, dephasing, addressability, and readout. A hydrogenic effective-mass estimate gives a characteristic acceptor localization length of order a few nanometers in Ge. Even allowing for central-cell corrections and valence-band anisotropy, this length scale remains far smaller than the proposed mean acceptor spacing of \(d_{\mathrm{avg}}\sim 170\,\mathrm{nm}\). Equivalently, the dimensionless overlap parameter \(n_{\mathrm{In}}^{1/3}a_B^*\) is well below the Mott limit for a plausible \(a_B^*\sim 2\)--\(5\,\mathrm{nm}\), supporting the assumption that the proposed density remains in the isolated-acceptor regime rather than forming a conducting impurity band. This also means that direct tunnel or exchange coupling at the mean spacing is expected to be weak; such coupling would require selected closer pairs, approximately \(10\)--\(30\,\mathrm{nm}\), or strong gate-induced hybridization.

\begin{table}[htbp]
\centering
\small
\caption{Comparison of common group-III acceptors in Ge. The ionization energies are shallow and similar for B, Al, Ga, and In, so the choice of In is motivated mainly by controlled incorporation, calibratable segregation behavior, and compatibility with the proposed statistically defined acceptor-chain architecture. The covalent-radius mismatch is estimated using \(r_{\mathrm{Ge}}\approx 120\,\mathrm{pm}\).}
\label{tab:groupIII_acceptors_ge}
\begin{tabularx}{\textwidth}{l c c c X X}
\toprule
\textbf{Acceptor} & \(E_A\) (meV) & \(r_i\) (pm) & \(\eta_i\) & \textbf{Growth behavior} & \textbf{Relevance for qubits} \\
\midrule
B  & \(\sim 10.4\) & 84  & \(-0.30\)  & Common shallow acceptor; strong size mismatch & Shallow and well known, but the large radius mismatch can increase local strain effects. \\
Al & \(\sim 10.2\) & 121 & \(+0.008\) & Common residual acceptor in HPGe & Very small mismatch; more relevant here as a background impurity than as the preferred intentional qubit species. \\
Ga & \(\sim 10.8\) & 122 & \(+0.017\) & Common residual acceptor in HPGe & Shallow and low-strain, but often treated as a background impurity in detector-grade material. \\
In & \(\sim 11.2\) & 142 & \(+0.183\) & Strongly segregating; calibratable by growth position and wafer selection & Selected here as the intentional acceptor because it can define localized hole states while remaining dilute enough to avoid impurity-band formation. \\
\bottomrule
\end{tabularx}
\end{table}

For a target local indium density
\begin{equation}
n_{\mathrm{In}}
=
2\times10^{14}\ \mathrm{cm^{-3}},
\label{eq:nin}
\end{equation}
the corresponding mean three-dimensional spacing is approximately
\begin{equation}
d_{\mathrm{avg}}
\approx
n_{\mathrm{In}}^{-1/3}
\approx
1.7\times10^{-5}\ \mathrm{cm}
\approx
170\ \mathrm{nm}.
\label{eq:davg}
\end{equation}
If the electrostatically active transverse mode area is chosen to be of order \((150\text{--}170~\mathrm{nm})^2\), a \SI{1}{\micro\meter}-long quasi-one-dimensional channel contains on the order of five acceptors on average at this density. This is precisely the regime of interest for a statistically selected five-qubit proof-of-principle register or a short quantum bus. Importantly, this mesoscopic length scale is also compatible with phononic-crystal design, since submicron acceptor arrays and GHz-range acoustic cavities naturally occupy the same device-scale regime.

At the proposed density, the device is intended to operate in the freeze-out regime, where holes remain localized on individual acceptors rather than forming a conducting impurity band. This assumption must be verified experimentally through low-temperature transport, leakage-current measurements, and charge-sensing spectroscopy.

\subsection{Residual-strain argument for detector-grade Ge}

The preceding discussion makes clear that the qubit is intentionally strain-sensitive, so it is essential to distinguish \emph{controlled} strain engineering from \emph{uncontrolled} parasitic strain. For acceptor qubits, the relevant materials question is not only average impurity density but also whether impurity-induced strain is small enough that it does not dominate the heavy-hole/light-hole splitting or randomize the qubit spectrum. A simple order-of-magnitude estimate can be obtained from a Vegard-like mismatch model. If $N_0$ is the atomic density of Ge,
\begin{equation}
N_0
=
\frac{8}{a_0^3}
\approx
4.42\times10^{22}\ \mathrm{cm^{-3}},
\qquad
a_0 \approx 5.658~\text{\AA},
\label{eq:N0}
\end{equation}
then the average impurity-induced linear strain from species $i$ may be estimated as
\begin{equation}
\bar{\epsilon}_i
\sim
\eta_i\,\frac{n_i}{N_0},
\qquad
\eta_i
=
\frac{r_i-r_{\mathrm{Ge}}}{r_{\mathrm{Ge}}},
\label{eq:strain_est}
\end{equation}
where $r_i$ and $r_{\mathrm{Ge}}$ are covalent radii \cite{Cordero2008}. Using $r_{\mathrm{In}}\approx 142~\mathrm{pm}$ and $r_{\mathrm{Ge}}\approx 120~\mathrm{pm}$ gives
\begin{equation}
\eta_{\mathrm{In}}
\approx
\frac{142-120}{120}
\approx
0.183.
\end{equation}
At the target indium density of Eq.~\eqref{eq:nin},
\begin{equation}
\bar{\epsilon}_{\mathrm{In}}
\sim
0.183\times\frac{2\times10^{14}}{4.42\times10^{22}}
\approx
8\times10^{-10},
\label{eq:in_strain_value}
\end{equation}
which is far below $10^{-7}$. For the residual background impurities of detector-grade Ge,
\begin{equation}
n_{\mathrm{bg}}
\sim
10^{10}\ \mathrm{cm^{-3}}.
\label{eq:nbg}
\end{equation}
the corresponding average strain is smaller still. Thus, at the level of average impurity-induced hydrostatic strain, detector-grade Ge with controlled In doping should remain in a regime where unintended strain is negligible compared with the intentionally engineered electric, magnetic, and phononic control fields.

This estimate should be interpreted as a design-level scaling argument rather than a replacement for full atomistic modeling. In a real device, the more important quantities are the \emph{fluctuating} mesoscopic strain field over the acceptor wavefunction and the local symmetry breaking induced by nearby interfaces and electrodes. Nevertheless, Eqs.~\eqref{eq:strain_est}--\eqref{eq:in_strain_value} show why detector-grade Ge is such a favorable host: the deliberate acceptor array can set the qubit architecture while the uncontrolled impurity background remains parametrically too weak to dominate the acceptor-hole physics. This low-disorder starting point is also what makes phononic cavity engineering meaningful: the cavity can then be designed to control the relevant strain field rather than merely compensate for uncontrolled bulk disorder.

\subsection{Design implication}

Taken together, the physics arguments of this section support a coherent five-qubit design window. Detector-grade Ge supplies a low-disorder, weak-hyperfine host; In acceptors provide atom-like confinement; interface asymmetry and gate fields split the acceptor manifold into an electrically controllable qubit doublet; and the target density \(n_{\mathrm{In}}\sim 2\times 10^{14}\,\mathrm{cm^{-3}}\) naturally yields an active-volume regime in which a 1~\(\mu\mathrm{m}\)-long channel can contain about five acceptors on average.
Within this picture, a phononic crystal cavity is best interpreted as an engineered enhancement layer: it can suppress unwanted environmental phonons and preserve selected acoustic modes, but the baseline qubit concept does not depend on it. The platform therefore occupies a useful middle ground between donor qubits and gate-defined hole dots, combining chemistry-defined localization with the all-electrical controllability of Ge hole systems.

\section{Strain and Disorder Analysis}
\label{sec:strain_disorder}

Strain and disorder are central design variables for a Ge acceptor-spin platform because the low-energy spectrum of an acceptor-bound hole is highly sensitive to local symmetry breaking, heavy-hole/light-hole (HH/LH) mixing, and electric-field inhomogeneity \cite{AbadilloUriel2016,Scappucci2021}. In contrast to gate-defined hole qubits, where heterostructure strain is often deliberately used to engineer the valence-band manifold, the present platform seeks to exploit detector-grade bulk Ge precisely because it minimizes uncontrolled background strain and disorder. The design goal is therefore not to eliminate all strain, but to ensure that parasitic strain remains far below the level of intentionally applied electric- and interface-driven tuning.

For the proposed device, it is useful to distinguish three conceptually different contributions:
\begin{enumerate}
    \item \textbf{Intentional qubit-defining acceptors}, namely the In dopants introduced at
    \[
    n_{\mathrm{In}} \sim 2\times 10^{14}~\mathrm{cm^{-3}},
    \]
    which are part of the designed qubit architecture rather than unwanted disorder.
    \item \textbf{Residual background impurities}, whose total density in detector-grade Ge is taken to be
    \[
    n_{\mathrm{bg}} \sim 10^{10}~\mathrm{cm^{-3}},
    \]
    and which provide the dominant uncontrolled bulk contribution to electrostatic and strain disorder.
    \item \textbf{Interface- and gate-induced disorder}, including dielectric charge traps, local electric-field nonuniformity, and strain generated by the gate stack or surface processing.
\end{enumerate}

This hierarchy is important for the rest of the paper. The first category defines the qubit array itself, the second determines the residual bulk disorder floor, and the third is expected to dominate the remaining device-to-device variability once detector-grade purity is achieved. The present section therefore focuses on the bulk-material contribution from the first two terms and treats the interface contribution qualitatively.

\subsection{Impurity-induced strain background}

A simple and physically transparent estimate of the average strain caused by dilute substitutional impurities can be obtained from a defect-relaxation-volume picture. If impurity species $i$ changes the local atomic volume by $\Delta \Omega_i$, then the average volumetric strain is
\begin{equation}
\langle \mathrm{Tr}\,\epsilon \rangle_i \simeq c_i \frac{\Delta \Omega_i}{\Omega_0},
\label{eq:volumetric_strain}
\end{equation}
where $c_i = n_i/N_0$ is the atomic fraction of species $i$, $n_i$ is its number density, and $N_0$ is the atomic density of Ge. For diamond-cubic Ge,
\begin{equation}
N_0 = \frac{8}{a_0^3} \approx 4.42\times 10^{22}~\mathrm{cm^{-3}},
\qquad
a_0 \approx 5.658~\text{\AA}.
\label{eq:atomic_density}
\end{equation}
If the relaxation is approximately isotropic, the corresponding average linear strain is
\begin{equation}
\bar{\epsilon}_i \simeq \frac{1}{3}\langle \mathrm{Tr}\,\epsilon \rangle_i.
\label{eq:linear_strain}
\end{equation}

In the dilute limit, it is convenient to approximate the relaxation volume by the covalent-radius mismatch,
\begin{equation}
\frac{\Delta \Omega_i}{\Omega_0} \approx 3\eta_i,
\qquad
\eta_i \equiv \frac{r_i-r_{\mathrm{Ge}}}{r_{\mathrm{Ge}}},
\label{eq:mismatch_parameter}
\end{equation}
which yields the order-of-magnitude estimate
\begin{equation}
\bar{\epsilon}_i \approx \eta_i \frac{n_i}{N_0}.
\label{eq:average_strain_estimate}
\end{equation}
Equation~\eqref{eq:average_strain_estimate} is not intended as an atomistic prediction for a specific device, but it provides a useful comparative metric for assessing whether residual impurity strain is likely to be relevant at the qubit level.

For indium, using the covalent radii $r_{\mathrm{In}}\approx 142~\mathrm{pm}$ and $r_{\mathrm{Ge}}\approx 120~\mathrm{pm}$ \cite{Cordero2008}, we obtain
\begin{equation}
\eta_{\mathrm{In}} \approx \frac{142-120}{120} \approx 0.183.
\label{eq:eta_in}
\end{equation}
For the proposed intentional In density defined in Eq.~\eqref{eq:nin}, Eq.~\eqref{eq:average_strain_estimate} gives an average hydrostatic strain contribution of
\begin{equation}
\bar{\epsilon}_{\mathrm{In}}
\approx
0.183\times \frac{2\times 10^{14}}{4.42\times 10^{22}}
\approx
8.3\times 10^{-10}.
\label{eq:strain_in}
\end{equation}
Thus, even the intentional acceptor density contributes an average strain that is far below the $10^{-7}$ scale.

For the detector-grade residual background, the average strain is even smaller. Taking the residual concentration from Eq.~\eqref{eq:nbg} and using a conservative worst-case mismatch parameter $|\eta|\sim 0.3$ for an unintended impurity species, one finds
\begin{equation}
\bar{\epsilon}_{\mathrm{bg}}
\sim
0.3\times \frac{10^{10}}{4.42\times 10^{22}}
\approx
6.8\times 10^{-14}.
\label{eq:strain_bg}
\end{equation}
This is many orders of magnitude smaller than the strain levels typically used intentionally in strained Ge/SiGe heterostructures, where the relevant strain scale is often in the range $10^{-3}$--$10^{-2}$ \cite{Scappucci2021}. In that sense, detector-grade bulk Ge provides an essentially unstrained host on the scale relevant to parasitic impurity backgrounds.

\subsection{Mesoscopic strain fluctuations}

The mean strain alone is not the most relevant quantity for qubit reproducibility. What matters more in practice is the spatially averaged fluctuation over the finite volume sampled by the acceptor-bound hole. If the residual impurities are randomly distributed, the root-mean-square strain fluctuation over a volume $V$ scales as
\begin{equation}
\epsilon_{\mathrm{rms}}(V)
\approx
\left[
\sum_i
\frac{\eta_i^2 n_i}{N_0^2 V}
\right]^{1/2}.
\label{eq:rms_strain}
\end{equation}
This expression makes two important points explicit. First, the parasitic strain background decreases directly with improved chemical purity. Second, the fluctuation level is further reduced by spatial averaging over the finite extent of the acceptor-bound hole.

For a characteristic acceptor-bound-hole averaging volume set by several effective Bohr radii,
\begin{equation}
V \sim (10~\mathrm{nm})^3 = 10^{-18}~\mathrm{cm^3},
\label{eq:qubit_volume}
\end{equation}
and again using a conservative worst-case mismatch parameter $|\eta|\sim 0.3$ for the residual background, Eq.~\eqref{eq:rms_strain} gives
\begin{equation}
\epsilon_{\mathrm{rms}}^{(\mathrm{bg})}
\sim
\left[
\frac{(0.3)^2(10^{10})}{(4.42\times10^{22})^2(10^{-18})}
\right]^{1/2}
\approx
6.8\times10^{-10}.
\label{eq:rms_bg_value}
\end{equation}
This value is still well below $10^{-7}$ and supports the central design claim of the present architecture: in detector-grade Ge, the uncontrolled bulk strain background should be too small to dominate the acceptor-hole spectrum.

It is important, however, to distinguish the \emph{intentional} In dopants from the \emph{unintended} residual background. The In atoms define the qubit sites themselves and therefore are not disorder in the same sense as parasitic impurities. Their role is architectural. By contrast, the unwanted background impurities and dislocations are the sources of uncontrolled inhomogeneity. This distinction is one of the main reasons why detector-grade Ge is attractive: once the background is driven into the $10^{10}~\mathrm{cm^{-3}}$ regime, the dominant remaining sources of device-to-device variation are expected to arise from interfaces, gate fields, and local processing rather than from the bulk host crystal.

\subsection{Projected impact on the qubit Hamiltonian}

At the qubit level, strain and disorder enter primarily by shifting the acceptor energy levels, modifying HH/LH mixing, and perturbing the effective $g$ tensor and electric-dipole matrix elements. To leading order, the resulting qubit-frequency fluctuation may be written as
\begin{equation}
\delta\omega_q
\simeq
\frac{\partial \omega_q}{\partial \epsilon_{\mathrm{hyd}}}\,\delta\epsilon_{\mathrm{hyd}}
+
\frac{\partial \omega_q}{\partial \epsilon_{\mathrm{sh}}}\,\delta\epsilon_{\mathrm{sh}}
+
\frac{\partial \omega_q}{\partial E}\,\delta E,
\label{eq:qubit_fluctuation}
\end{equation}
where $\delta\epsilon_{\mathrm{hyd}}$ and $\delta\epsilon_{\mathrm{sh}}$ denote hydrostatic and shear-strain fluctuations, respectively, and $\delta E$ denotes local electric-field disorder. For acceptor-bound holes, the shear and symmetry-breaking terms are often more important than the purely hydrostatic shift because they directly perturb the spin-$3/2$ manifold and the HH/LH admixture \cite{AbadilloUriel2016}. Equation~\eqref{eq:qubit_fluctuation} therefore highlights the main design implication: once the bulk impurity background is suppressed to detector-grade levels, the dominant residual noise sources are expected to be interface disorder and local field inhomogeneity rather than average hydrostatic strain.

This point also clarifies where phononic-crystal engineering belongs in the overall paper. A phononic crystal cavity primarily reshapes the \emph{dynamic} acoustic environment and the phonon density of states seen by the qubit; it does not remove the \emph{static} impurity- and interface-induced strain floor quantified in the present section. For that reason, a full discussion of phononic-crystal cavities is more naturally deferred to the spin--phonon coupling and architecture sections, where it can be introduced as a tool for suppressing unwanted relaxation channels and engineering useful cavity-mediated interactions, rather than as a remedy for static bulk disorder.

\subsection{Design implications for the five-qubit In-acceptor platform}

The quantitative message of the above estimates is straightforward. At the proposed intentional In density, the mean hydrostatic strain contribution remains at the $\sim 10^{-9}$ level [Eq.~\eqref{eq:strain_in}], while the uncontrolled detector-grade background contributes only $\sim 10^{-14}$ to the mean strain [Eq.~\eqref{eq:strain_bg}] and $\sim 10^{-10}$ to the mesoscopic rms fluctuation [Eq.~\eqref{eq:rms_bg_value}]. These numbers are far below the $10^{-7}$ scale and many orders of magnitude below the engineered heterostructure-strain scales commonly used in Ge/SiGe hole devices. The corresponding order-of-magnitude strain scales are collected in Table~\ref{tab:strain_scales}.

Accordingly, the proposed platform should be viewed as operating in a regime where bulk impurity strain is not the dominant obstacle. Instead, the main disorder challenges are likely to be: (i) local interface symmetry breaking near the gate stack, (ii) electric-field nonuniformity from dielectric or surface traps, and (iii) any residual crystal defects or dislocation-related strain hotspots not captured by the dilute-impurity estimate. Even with these caveats, detector-grade Ge offers a uniquely favorable starting point for acceptor-spin qubits because it suppresses the uncontrolled bulk contribution to a level where intentional gate tuning, rather than parasitic material disorder, should dominate the qubit Hamiltonian.

This conclusion is important for the coherence of the full manuscript. The role of the present section is to show that detector-grade Ge can suppress the static strain/disorder floor to a level compatible with acceptor qubits. The role of the later spin--phonon and device-architecture sections is then to show how the remaining dynamic acoustic environment may be engineered---for example, with a phononic crystal cavity---to further improve relaxation, coupling selectivity, and architectural scalability. The two ideas are complementary, but they solve different problems.

\begin{table}[H]
\centering
\caption{Order-of-magnitude strain scales relevant to the proposed five-qubit In-acceptor platform in detector-grade Ge. The quoted values are design-level estimates intended to compare intentional and parasitic strain sources rather than to replace atomistic device simulation.}
\label{tab:strain_scales}
\begin{tabular}{lccc}
\hline
Source & Density ($\mathrm{cm^{-3}}$) & Representative $|\eta|$ & Estimated strain scale \\
\hline
Intentional In acceptors & $2\times10^{14}$ & $0.183$ & $\bar{\epsilon}_{\mathrm{In}}\approx 8.3\times10^{-10}$ \\
Residual background (13N) & $10^{10}$ & $0.3$ & $\bar{\epsilon}_{\mathrm{bg}}\approx 6.8\times10^{-14}$ \\
Residual background over $(10~\mathrm{nm})^3$ & $10^{10}$ & $0.3$ & $\epsilon_{\mathrm{rms}}\approx 6.8\times10^{-10}$ \\
Typical strained Ge/SiGe heterostructure & --- & --- & $10^{-3}$--$10^{-2}$ \\
\hline
\end{tabular}
\end{table}

\section{Spin--phonon Coupling and Coupling Mechanisms}
\label{sec:spin_phonon}
\label{sec:spin_phonon_coupling}

For an In acceptor qubit in detector-grade Ge, phonons are not merely a secondary decoherence channel; they are part of the control and coupling landscape of the qubit itself. This follows from the spin-$3/2$ valence-band character of the acceptor-bound hole: acoustic strain acts directly on the heavy-hole/light-hole (HH/LH) manifold, modifies orbital admixture, and thereby changes both the effective $g$ tensor and the electric-dipole matrix elements relevant for qubit manipulation \cite{AbadilloUriel2016,Terrazos2021,Wang2024Modeling}. In this sense, an acceptor qubit in Ge is intrinsically both spin--orbit-active and strain-active, making it a natural candidate for all-electrical control, strain-assisted driving, and eventually phonon-mediated coupling.

At the same time, the same strain sensitivity that makes the qubit controllable also exposes it to acoustic relaxation and dephasing. This creates an architectural choice. A first-generation five-qubit chain may be realized without explicit phononic engineering, relying only on local electrical control and the naturally low bulk disorder of detector-grade Ge. However, once one seeks improved relaxation protection or nonlocal coupling beyond nearest-neighbor dipole and exchange channels, the local acoustic environment becomes an important design parameter. For that reason, a phononic crystal cavity (PnC) should be viewed not as a mandatory starting requirement, but as a highly attractive next-stage enhancement of the present platform \cite{Smelyanskiy2014GePhononic,Mei2025QST}.

\subsection{Direct phonon--charge coupling versus effective phonon--spin coupling}

It is useful to distinguish two related but physically distinct couplings. The first is the \emph{direct phonon--charge} (or more precisely phonon--orbital / strain--orbital) coupling, which describes how lattice deformation perturbs the orbital structure of the localized acceptor-bound hole. The second is the \emph{effective phonon--spin} coupling, which arises only after the strain-induced orbital response is converted into qubit rotations or qubit-frequency shifts through spin--orbit interaction and HH/LH mixing.

For acceptor-bound holes in Ge, a compact starting form is
\begin{equation}
H_{\mathrm{ph}}
=
H_{\mathrm{BP}}\!\left[\epsilon_{\mathrm{ph}}(\mathbf{r},t)\right]
+
H_{\mathrm{int}},
\label{eq:Hph_total}
\end{equation}
where $H_{\mathrm{BP}}$ is the Bir--Pikus strain Hamiltonian acting on the valence-band manifold and $H_{\mathrm{int}}$ represents interface- and confinement-induced symmetry breaking. After projection into the lowest acceptor Kramers doublet, the resulting qubit-level coupling may be written schematically as
\begin{equation}
H_{q\text{-}\mathrm{ph}}
\approx
\frac{\mu_B}{2}\,
\mathbf{B}\!\cdot\!
\left(
\frac{\partial \mathbf{g}}{\partial \epsilon}
:
\epsilon_{\mathrm{ph}}
\right)
\!\cdot\!
\bm{\sigma}
+
H_{\mathrm{mix}}\!\left[\epsilon_{\mathrm{ph}}\right],
\label{eq:Hqph_general}
\end{equation}
where the first term describes strain-dependent modulation of the effective $g$ tensor and the second term collects strain-driven orbital and HH/LH mixing processes that convert a nominally spin-like qubit into a spin--orbit-active two-level system \cite{Terrazos2021,AbadilloUriel2023StrainSOI}. Equation~\eqref{eq:Hqph_general} captures the key distinction of Ge acceptor and hole qubits relative to simpler electron-spin systems: strain does not merely perturb the orbital sector weakly and indirectly, but acts directly on the spin--orbit-active valence manifold.

A useful local linearization of Eq.~\eqref{eq:Hqph_general} around a chosen operating point is
\begin{equation}
H_{q\text{-}\mathrm{ph}}
=
\frac{1}{2}
\sum_{\alpha=x,y,z}
\lambda_{\alpha}\,
\epsilon_{\alpha}^{\mathrm{ph}}(t)\,
\sigma_{\alpha},
\label{eq:Hqph_linear}
\end{equation}
where the coefficients $\lambda_{\alpha}$ summarize the effective coupling of the relevant strain channels to the qubit Pauli operators. In practice, the dominant channels depend on the magnetic-field orientation, acceptor depth, interface asymmetry, and gate bias. This effective form also makes clear why phononic engineering is meaningful: once the strain field seen by the qubit is structured, the qubit response can be structured with it.

\subsection{Why a phononic crystal cavity is a natural extension}

The most direct motivation for adding a phononic crystal cavity is that it addresses the central trade-off of Ge hole-based qubits: the qubit benefits from strong strain sensitivity for control, but that same sensitivity opens unwanted acoustic relaxation channels. At lowest order, the phonon-limited relaxation rate may be written schematically as
\begin{equation}
\Gamma_{1}^{\mathrm{ph}}
\propto
|g_{\mathrm{sp}}|^2 \rho_{\mathrm{ph}}(\omega_q),
\label{eq:Gamma1_DOS}
\end{equation}
where $g_{\mathrm{sp}}$ is the effective spin--phonon matrix element and $\rho_{\mathrm{ph}}(\omega_q)$ is the local phononic density of states at the qubit transition frequency $\omega_q$. In bulk material, $\rho_{\mathrm{ph}}(\omega_q)$ is generally nonzero, so one-phonon relaxation channels remain open. By contrast, if $\omega_q$ lies inside a phononic bandgap,
\begin{equation}
\rho_{\mathrm{ph}}(\omega_q)\rightarrow 0
\qquad
\text{(inside the bandgap)},
\label{eq:pnc_bandgap}
\end{equation}
then the corresponding unwanted relaxation channel is strongly suppressed \cite{Smelyanskiy2014GePhononic,Mei2025QST}.

This means that a PnC cavity can improve performance in two complementary ways. First, it can act as an \emph{acoustic shield}, reducing environmental phonons at the qubit frequency and thereby improving $T_1$. Second, it can act as a \emph{mode selector}: by introducing a defect cavity or guided channel into the phononic crystal, one may preserve only those acoustic modes that are useful for coherent control or mediated coupling. In other words, the PnC allows one to suppress unwanted phonons without suppressing all phonons.

A minimal cavity-enabled Hamiltonian is
\begin{equation}
H_{\mathrm{cav}}
=
\hbar \omega_{\mathrm{ph}} a^\dagger a
+
\sum_i \frac{\hbar \omega_{q,i}}{2}\sigma_i^z
+
\sum_i \hbar g_i (a+a^\dagger)\sigma_i^x,
\label{eq:Hcav}
\end{equation}
where $a^\dagger$ and $a$ create and annihilate a localized phononic cavity mode of frequency $\omega_{\mathrm{ph}}$, and $g_i$ is the spin--phonon coupling rate for qubit $i$ in angular-frequency units. The explicit factor $\hbar$ in the Hamiltonian converts this rate to an energy, while quoted values such as $g_i/2\pi$ are reported in hertz. At the design level, the coupling strength can be related to the zero-point strain field of the cavity mode,
\begin{equation}
g_i
\sim
\frac{1}{2\hbar}
\sum_{\alpha}
\lambda_{\alpha}\,
\epsilon_{\alpha,i}^{\mathrm{zpf}},
\label{eq:g_zpf}
\end{equation}
where $\epsilon_{\alpha,i}^{\mathrm{zpf}}$ is the zero-point strain projected onto the relevant qubit-coupling channels. For Ge hole-spin systems, recent phononic-cavity modeling has reported representative spin--phonon coupling strengths up to
\begin{equation}
g/2\pi \sim 6.3~\mathrm{MHz},
\end{equation}
together with cavity quality factors
\begin{equation}
Q_{\mathrm{ph}} > 10^{4},
\end{equation}
and phonon-mediated $T_1$ values reaching the millisecond scale when the acoustic density of states is appropriately engineered \cite{Mei2025QST}. These values are not yet specific benchmarks for the present In-acceptor chain, but they show that a cavity-protected and cavity-coupled operating regime is realistic for Ge hole-based qubits.

For the coherence of the present paper, the architectural conclusion is therefore the following: a PnC cavity is \emph{not} required to define or operate the first five-qubit In-acceptor chain, but it becomes highly valuable once one seeks either improved phonon-limited relaxation or a purposeful nonlocal coupling channel. It should therefore be treated as a natural second-stage enhancement of the platform rather than as a prerequisite for the basic five-qubit demonstration.

\subsection{Single-qubit strain driving}

With this role of the acoustic environment established, one can return to the single-qubit control problem. Because the acceptor qubit possesses both spin--orbit activity and an electric dipole, an oscillatory strain field can drive the qubit in close analogy to electric-dipole spin resonance (EDSR). If a coherent acoustic mode generates a time-dependent strain field
\begin{equation}
\epsilon_{\alpha}^{\mathrm{ph}}(t)
=
\epsilon_{\alpha}^{(0)}
\cos(\omega t),
\end{equation}
then the corresponding strain-driven Rabi frequency is
\begin{equation}
\Omega_{\mathrm{ph}}
=
\frac{2}{\hbar}
\left|
\left\langle 1 \left| H_{q\text{-}\mathrm{ph}}^{(\mathrm{ac})} \right| 0 \right\rangle
\right|,
\label{eq:Omega_ph}
\end{equation}
where $|0\rangle$ and $|1\rangle$ are the qubit states. This expression makes clear that the strength of phonon-driven control depends directly on how efficiently the oscillatory strain modulates the HH/LH admixture, the effective Zeeman tensor, or the electric-dipole-active excited-doublet structure of the acceptor.

The literature on Ge hole-spin qubits suggests that this strain sensitivity can be substantial. In particular, strain gradients may strongly enhance hole-spin driving by generating additional Rashba-like spin--orbit terms and $g$-tensor modulations. For Ge-based hole systems, shear-strain gradients as small as
\begin{equation}
\partial_x \epsilon_{yz} \sim 3\times10^{-6}~\mathrm{nm^{-1}}
\end{equation}
have been shown to enhance Rabi frequencies by roughly one order of magnitude \cite{AbadilloUriel2023StrainSOI}. Although that result was derived for Ge/GeSi hole quantum dots rather than substitutional In acceptors, the underlying lesson carries over directly: in Ge hole-based qubits, strain is not simply an unwanted perturbation but a potentially powerful control resource. In a later-generation implementation, a PnC cavity could further enhance this mechanism by concentrating the strain field into a designed cavity mode rather than relying only on extended or weakly confined acoustic excitations.

\subsection{Coupling mechanisms between neighboring acceptor qubits}

For the present five-qubit architecture, several coupling mechanisms are in principle available. The most relevant ones are dipole--dipole coupling, short-range exchange coupling, and phonon-mediated coupling through localized or guided acoustic modes. The first two do not require explicit phononic engineering; the third becomes especially attractive once a PnC cavity or waveguide is added.

\subsubsection{Electric dipole--dipole coupling}

If the acceptor-bound hole is displaced relative to the ionized In core by an amount $\Delta \mathbf{r}_i$, then the local dipole moment of qubit $i$ is
\begin{equation}
\mathbf{p}_i = e\,\Delta \mathbf{r}_i.
\label{eq:dipole_moment}
\end{equation}
For two acceptors separated by $\mathbf{r}_{ij}=r_{ij}\hat{\mathbf{r}}_{ij}$, the electrostatic dipole--dipole interaction is described in Eq.~\eqref{eq:dipole_dipole}.

This interaction is attractive architecturally because it is long-ranged compared with direct wavefunction overlap and can be tuned electrically through the acceptor dipole moments. In a one-dimensional chain with average spacing $d_{\mathrm{avg}}\sim \SI{170}{\nano\meter}$, Eq.~\eqref{eq:dipole_dipole} suggests that nearest-neighbor electrostatic coupling can be useful for favorable dipole orientations and sufficient gate-induced dipole moments, without requiring direct wavefunction overlap to dominate.

\subsubsection{Exchange coupling}

If neighboring acceptor wavefunctions overlap sufficiently, a short-range exchange channel also becomes available. Because the relevant bound-state length scale is only \(a_B^*\sim 2\)--\(5\,\mathrm{nm}\), this mechanism is not expected to be strong for typical \(\sim\SI{170}{\nano\meter}\) separations; it is relevant mainly for unusually close statistically selected pairs, roughly \(10\)--\(30\,\mathrm{nm}\), or for configurations in which gate fields deliberately increase hybridization. At the simplest Hubbard-like level, the effective exchange may be written as
\begin{equation}
J_{ij}^{\mathrm{ex}} \sim \frac{4 t_{ij}^{2}}{U},
\label{eq:Jex}
\end{equation}
where $t_{ij}$ is the tunnel amplitude between neighboring acceptor orbitals and $U$ is an effective on-site charging energy. In practice, $t_{ij}$ depends exponentially on acceptor spacing, acceptor depth, and local electric fields. For this reason, exchange is potentially powerful for nearest-neighbor entangling gates, but it is also the most fabrication-sensitive channel among the mechanisms considered here.

\subsubsection{Phonon-mediated coupling and the role of a PnC cavity}

The distinctive opportunity of a Ge acceptor platform is that the same strain sensitivity that enables fast single-qubit driving can also be used to couple distant qubits through a shared acoustic mode. For a cavity or guided phonon mode of frequency $\omega_{\mathrm{ph}}$, a minimal qubit--phonon Hamiltonian is
\begin{equation}
H_{\mathrm{bus}}
=
\hbar \omega_{\mathrm{ph}} a^{\dagger}a
+
\sum_i
\frac{\hbar \omega_{q,i}}{2}\sigma_i^z
+
\sum_i
\hbar g_i
\left(a+a^{\dagger}\right)\sigma_i^x,
\label{eq:Hbus}
\end{equation}
where $a^{\dagger}$ and $a$ create and annihilate the relevant phonon mode and $g_i$ is the single-qubit spin--phonon coupling rate in angular-frequency units. With this convention, $g_i/2\pi$ is quoted in megahertz and the Hamiltonian contains $\hbar g_i$; equivalently, the same expression can be written for $H/\hbar$ without the explicit $\hbar$. In the dispersive regime, where the qubits are detuned from the phonon mode by $\Delta_i=\omega_{q,i}-\omega_{\mathrm{ph}}$, the phonon can mediate an effective inter-qubit interaction rate of order
\begin{equation}
J_{ij}^{\mathrm{ph}}
\sim
\frac{g_i g_j}{\Delta},
\label{eq:Jph}
\end{equation}
for comparable detunings $\Delta_i\approx \Delta_j\approx \Delta$. This phonon-bus picture is attractive because it can extend the interaction range beyond nearest-neighbor electrostatics or exchange and may eventually enable modular coupling between multiple qubits or subregisters.

Here the benefit of a PnC cavity is especially clear. Without phononic structuring, phonon-mediated coupling competes with a dense continuum of unwanted acoustic modes. With a PnC cavity or waveguide, the phonon spectrum can be narrowed to one or a few selected modes inside an otherwise gapped acoustic environment. The cavity therefore improves the selectivity of the phonon bus: it not only supplies the relevant mode, but suppresses the surrounding unwanted modes that would otherwise contribute to decoherence and crosstalk. In that sense, the PnC cavity is not merely a stronger coupler; it is a spectral filter that makes coherent coupling more controllable.

\subsection{Competition between control and relaxation}

The same coupling that enables phonon-assisted control can also drive unwanted relaxation. At lowest order, the phonon-limited relaxation rate may be written schematically as
\begin{equation}
\Gamma_{1}^{\mathrm{ph}}
=
\frac{2\pi}{\hbar}
\sum_{\mathbf{q},\nu}
\left|
\left\langle 0;1_{\mathbf{q}\nu}\left|
H_{q\text{-}\mathrm{ph}}
\right|1;0\right\rangle
\right|^{2}
\delta\!\left(\hbar\omega_q-\hbar\omega_{\mathbf{q}\nu}\right),
\label{eq:T1goldenrule}
\end{equation}
where $\omega_q$ is the qubit splitting and $\omega_{\mathbf{q}\nu}$ is the phonon frequency of branch $\nu$ and wavevector $\mathbf{q}$. Equation~\eqref{eq:T1goldenrule} highlights the central design trade-off of the platform: stronger strain susceptibility improves controllability but also strengthens the qubit's coupling to the acoustic bath.

\subsection{Relaxation-limited coherence and the upper bound on $T_2$}

The phonon-limited relaxation rate in Eq.~\eqref{eq:T1goldenrule} also provides a useful way to
state the long-coherence potential of the platform. For a spin qubit, the
transverse coherence time is bounded by the longitudinal relaxation time through
the standard relation
\begin{equation}
\frac{1}{T_2}=\frac{1}{2T_1}+\frac{1}{T_\phi},
\label{eq:T2_relation}
\end{equation}
where $T_\phi$ is the pure-dephasing time associated with slow fluctuations in
the qubit frequency. Thus, in the ideal relaxation-limited regime,
\begin{equation}
T_{2,\mathrm{max}} = 2T_1
\qquad
\left(T_\phi^{-1}\rightarrow 0\right).
\label{eq:T2max}
\end{equation}
This relation is important for interpreting the role of detector-grade Ge and
phononic-crystal engineering. Detector-grade Ge is expected to suppress the
static bulk disorder contribution to qubit-frequency inhomogeneity, while
isotopic purification, low-noise gates, operation near electric- or magnetic-field
sweet spots, and improved interface control can reduce residual pure dephasing.
If these non-relaxation dephasing channels are sufficiently controlled, then the
remaining coherence limit is set primarily by phonon-mediated spin relaxation.

In this relaxation-limited regime, the benefit of a phononic crystal is not only
to increase $T_1$, but also to raise the possible ceiling for $T_2$. If the
phononic environment suppresses the relevant one-phonon density of states by a
factor $S_{\mathrm{PnC}}<1$, then, to leading order,
\begin{equation}
\Gamma_{1,\mathrm{PnC}}^{\mathrm{ph}}
\simeq S_{\mathrm{PnC}}\Gamma_{1,\mathrm{bulk}}^{\mathrm{ph}},
\qquad
T_{1,\mathrm{PnC}}^{\mathrm{ph}}
\simeq \frac{T_{1,\mathrm{bulk}}^{\mathrm{ph}}}{S_{\mathrm{PnC}}}.
\end{equation}
Consequently, a suppression factor in the range
$S_{\mathrm{PnC}}\sim 10^{-2}$--$10^{-3}$ could raise a millisecond-scale
phonon-limited relaxation time into the sub-second to seconds range, provided
that other dephasing mechanisms remain below the relaxation floor. Under those
conditions, the corresponding upper bound
$T_2 \leq 2T_1$ implies that seconds-level coherence is a conditional
long-term target for optimized Ge acceptor-spin devices. This estimate should be
interpreted as an upper-bound design target, not as a predicted device
performance or a first-generation benchmark.

This is precisely the point at which a PnC cavity changes the physics from a liability into a resource. In an unstructured acoustic environment, increasing the spin--phonon matrix element generally strengthens both the desired control channel and the unwanted relaxation channel. A PnC environment changes this balance by suppressing the broad continuum of environmental acoustic modes while preserving a small number of engineered cavity or guided modes. \textbf{The resulting design principle is strong coupling to selected phononic modes while suppressing coupling to the uncontrolled acoustic bath.} This is why phononic engineering is valuable not only as a coupling resource, but also as a coherence-protection strategy for extending the relaxation-limited ceiling of the qubit coherence time.

\subsection{Design implications for the five-qubit In-acceptor chain}

For the present five-qubit architecture, the practical interaction hierarchy is clear. Local electric-field tuning remains the main tool for single-qubit addressability because the In acceptor is intrinsically electric-dipole-active and spin--orbit-active. Nearest-neighbor dipole and exchange coupling provide the most natural short-range two-qubit channels, with dipole coupling expected to be more robust against fabrication variability and exchange potentially stronger but more sensitive to local structure. Phonon-mediated coupling is the most promising route to extend the interaction range beyond nearest neighbors and toward one-dimensional quantum-bus behavior within the same detector-grade Ge host.

Thus, the baseline five-qubit chain can first test electrical control, addressability, and nearest-neighbor coupling without explicit phononic structuring. A PnC branch is most useful in the next stage, when the goal shifts to improved phonon-limited relaxation, spectral filtering, and selected medium-range coupling through cavity or guided modes.

\section{Design Modeling Workflow}
\label{sec:design_modeling_workflow}

Because the proposed five-qubit platform spans multiple physical scales---from impurity incorporation during crystal growth, to gate-defined electrostatics, to acceptor-level spin physics, and finally to multi-qubit dynamics---its design requires a multiscale modeling workflow rather than a single monolithic model. The purpose of the workflow is to identify a self-consistent operating window in which atom-like localization, electrical addressability, inter-qubit coupling, and disorder tolerance can coexist within a \SI{1}{\micro\meter} detector-grade Ge channel. This section should therefore be read as a \emph{nominal feasibility modeling workflow}: it combines literature-grounded physical inputs, design-level calculations, and representative outputs for a self-consistent device concept, but it is not intended to replace a fully closed device-specific validation campaign.

To avoid overinterpreting the numerical values below, we define the assumptions used for the nominal estimates explicitly. Unless otherwise stated, the values reported in this section should be read as design estimates based on analytic scaling, literature-calibrated parameters, and simplified electrostatic and phononic models. They are intended to establish a plausible operating hierarchy, not to serve as final device-specific finite-element or atomistic benchmarks. A full validation study would require three-dimensional electrostatic modeling, atomistic or multiband acceptor modeling, disorder-ensemble sampling, and measured device parameters from fabricated structures. The nominal parameter set used for these estimates is listed in Table~\ref{tab:nominal_assumptions}.

\begin{table}[p]
\centering
\small
\caption{Nominal assumptions used for the design estimates in Section~6. These parameters define a representative operating window for the proposed five-qubit In-acceptor chain; they should be refined by finite-element electrostatics, multiband acceptor calculations, and experimental calibration in future device-specific studies.}
\label{tab:nominal_assumptions}
\begin{tabularx}{\textwidth}{p{4.0cm}X}
\toprule
\textbf{Quantity} & \textbf{Nominal assumption or design range} \\
\midrule
Active channel length & \(L=\SI{1}{\micro\meter}\) \\
Effective transverse active area & Order \((150\text{--}170\,\mathrm{nm})^2\), giving \(\sim 5\) acceptors on average at \(n_{\mathrm{In}}\sim 2\times10^{14}\,\mathrm{cm^{-3}}\) \\
Intentional In density & \(n_{\mathrm{In}}\sim 2\times10^{14}\,\mathrm{cm^{-3}}\) \\
Residual background density & \(n_{\mathrm{bg}}\sim 10^{10}\,\mathrm{cm^{-3}}\) \\
Mean acceptor spacing & \(d_{\mathrm{avg}}\sim 170\,\mathrm{nm}\) \\
Representative acceptor depth & Tens of nanometers below the gate dielectric; chosen to balance gate tunability against interface-induced noise \\
Gate pitch & Matched approximately to the acceptor spacing, typically \(\sim 150\)--\(200\,\mathrm{nm}\) \\
Gate dielectric thickness & Few-nanometer to \(\sim 10\,\mathrm{nm}\) scale, depending on leakage and interface-quality constraints \\
Dielectric constants & \(\varepsilon_{\mathrm{Ge}}\approx 16\); oxide/dielectric value chosen according to the gate-stack material \\
Gate-induced hole displacement & \(\Delta r\sim 0.5\)--\(1.0\,\mathrm{nm}\) for representative dipole-coupling estimates \\
Magnetic-field range & Chosen to place the qubit transition in the GHz regime while maintaining compatibility with EDSR and cryogenic wiring \\
Qubit-frequency range & GHz-scale transitions, with the precise value set by magnetic-field orientation, effective \(g\) tensor, strain, and gate bias \\
PnC target frequency & Matched to the selected qubit or cavity-coupling frequency; the PnC branch is treated as an engineered second-stage enhancement \\
Electrostatic model status & Nominal estimates based on simplified electrostatic scaling; final values require full three-dimensional finite-element modeling of the actual gate stack \\
Phononic model status & Nominal suppression and coupling estimates based on acoustic-density-of-states engineering; final values require device-specific PnC band-structure and cavity-mode modeling \\
\bottomrule
\end{tabularx}
\end{table}

For coherence with the rest of the paper, we distinguish between a \emph{baseline workflow} and an \emph{extended PnC-enhanced workflow}. The baseline workflow asks whether the five-qubit chain is viable using detector-grade Ge, intentional In acceptors, and local electrical control. The extended workflow then asks whether a phononic crystal cavity (PnC) provides enough additional relaxation protection or nonlocal coupling selectivity to justify its added fabrication complexity. In that sense, the PnC is treated as a second-stage architectural enhancement rather than as a prerequisite for the first-generation device.

The baseline workflow proceeds in five layers:
\begin{enumerate}
    \item \textbf{Materials and statistical layout generation}, which determines the expected acceptor positions, spacing statistics, and residual background disorder for a target indium concentration.
    \item \textbf{Electrostatic device modeling}, which computes the gate-induced potential, local electric fields, and site-to-site detuning under the plunger and barrier gates.
    \item \textbf{Single-acceptor quantum modeling}, which solves for the field-split acceptor manifold and extracts the effective qubit Hamiltonian for each site.
    \item \textbf{Inter-qubit coupling extraction}, which evaluates dipolar, exchange, and phonon-mediated interaction channels.
    \item \textbf{Chain-level dynamics and yield analysis}, which determines whether the five-site chain supports practical control, addressability, and coupling under realistic disorder.
\end{enumerate}
An engineered sixth step,
\begin{enumerate}
\setcounter{enumi}{5}
    \item \textbf{PnC cavity branch}, which evaluates whether an engineered acoustic bandgap and a selected defect-cavity mode improve performance enough to justify the added fabrication complexity.
\end{enumerate}

This layered structure is consistent with the broader Ge-hole literature, where the relevant physics emerges from the interplay of electrostatics, spin--orbit coupling, strain, and HH/LH mixing rather than from a purely isolated spin model \cite{Scappucci2021,Terrazos2021,Wang2024Modeling}. The results reported below should therefore be interpreted as \emph{nominal design estimates} for the nominal geometry, intended to establish physical plausibility and operating hierarchy rather than to claim final experimentally validated device benchmarks.

\subsection{Workflow overview}

\begin{figure}[htp!]
    \centering
    \includegraphics[width=\textwidth]{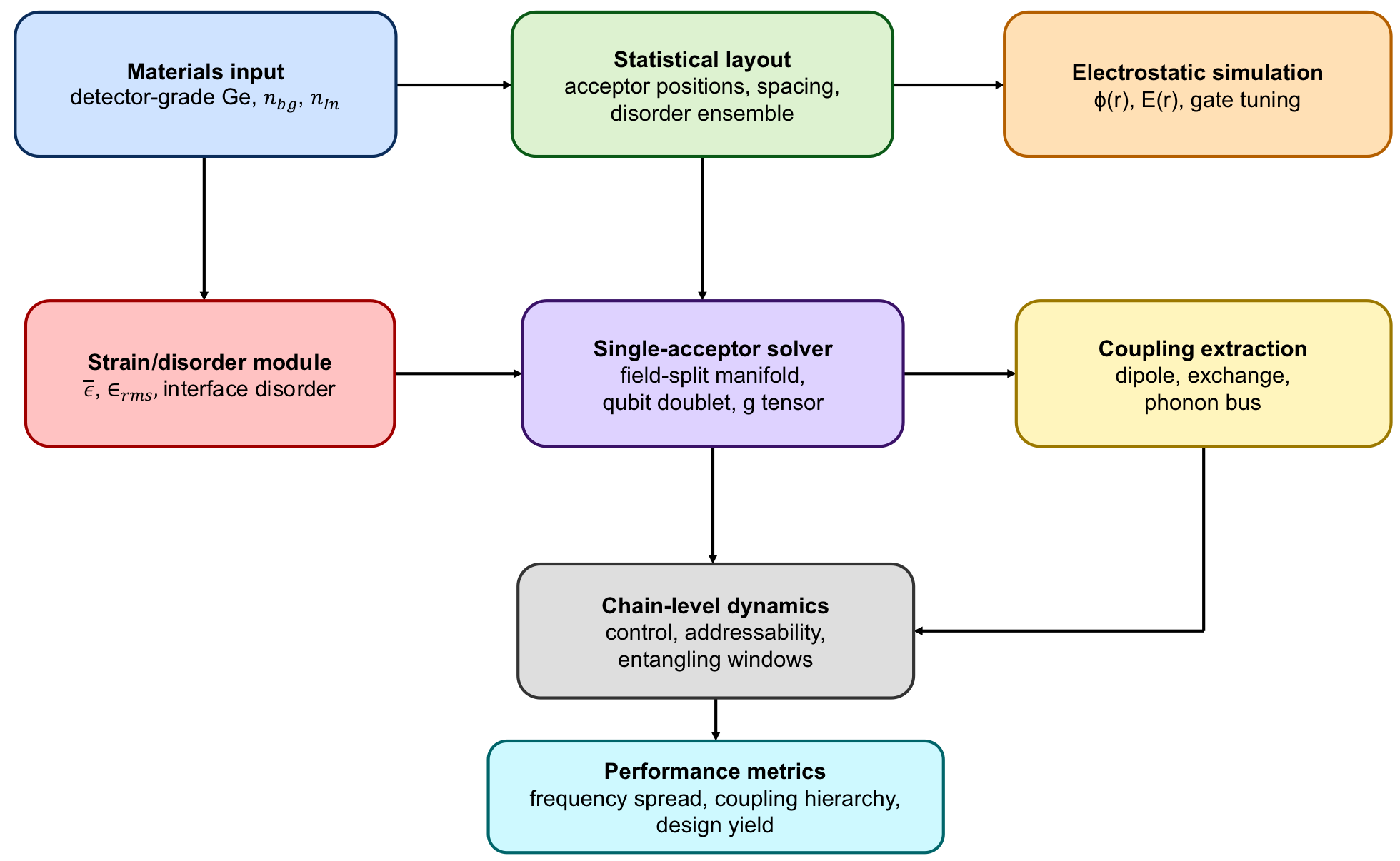}
    \caption{Multiscale modeling workflow for the proposed five-qubit In-acceptor platform in detector-grade Ge. The figure plots the logical data flow used in the design study: detector-grade materials inputs and a statistically generated In-acceptor layout feed electrostatic gate modeling, strain/disorder analysis, and single-acceptor Hamiltonian extraction; these outputs are then used to compute Stark tunability, qubit-frequency spread, pairwise dipolar/exchange-assisted coupling, and chain-level dynamical behavior. The optional phononic-crystal cavity branch plots how an engineered acoustic bandgap or defect mode modifies the local phonon density of states, the phonon-limited relaxation channel, and possible cavity-mediated coupling. The outputs plotted at the right of the workflow are the architecture-level performance metrics used to judge design feasibility, including addressability, coupling hierarchy, relaxation trends, and statistical device yield.}
    \label{fig:simulation_workflow}
\end{figure}

Figure~\ref{fig:simulation_workflow} summarizes the modeling workflow used for the present architecture. The central idea is to start from the material and device geometry, then progressively reduce the model to the qubit level while retaining the physically important quantities: local electric field, strain background, $g$-tensor anisotropy, dipole moment, and pairwise coupling strengths. In the extended branch, the same qubit parameters are then embedded into a phononic-cavity design space in order to determine whether acoustic band engineering provides a significant improvement in coherence or nonlocal coupling.

\subsection{Step 1: Materials and statistical layout generation}

The first step is to generate a statistical ensemble of acceptor configurations consistent with the target indium concentration in Eq.~\eqref{eq:nin}, embedded in detector-grade Ge with the residual background impurity concentration in Eq.~\eqref{eq:nbg}. At the nominal In density, the three-dimensional mean spacing is the value derived in Eq.~\eqref{eq:davg}.
A five-site register is obtained statistically by defining an active device volume rather than by assuming a perfectly deterministic one-dimensional line. For an active length \(L=\SI{1}{\micro\meter}\) and an electrostatically selected transverse area \(A_\perp\sim (150\text{--}170~\mathrm{nm})^2\), the mean number of intentional acceptors in the selected volume is
\begin{equation}
\bar{N}_{\mathrm{In}} = n_{\mathrm{In}} L A_\perp \sim 4.5\text{--}5.8,
\label{eq:Nbar_In_active}
\end{equation}
which is consistent with a statistically selected five-qubit architecture.
For a random dilute incorporation process, the simplest occupancy model is a
Poisson distribution,
\begin{equation}
P(N=k)=\frac{\bar{N}_{\mathrm{In}}^{\,k}e^{-\bar{N}_{\mathrm{In}}}}{k!}.
\label{eq:poisson_occupancy}
\end{equation}
For the nominal mean occupancy \(\bar{N}_{\mathrm{In}}=5\), this gives
\(P(N=5)\simeq 0.18\) and \(P(N\geq 5)\simeq 0.56\). Across the design
range in Eq.~\eqref{eq:Nbar_In_active}, the corresponding estimates are
\(P(N=5)\simeq 0.16\text{--}0.18\) and \(P(N\geq 5)\simeq 0.47\text{--}0.69\). Table~\ref{tab:poisson_occupancy} gives the corresponding occupancy distribution for the nominal case. These
numbers should be viewed as an upper bound on fabrication yield, because only
a subset of devices will have suitable spacing, acceptor depth, gate lever arms,
charge-sensor visibility, and coupling hierarchy.

\begin{table}[H]
\centering
\small
\caption{Poisson occupancy estimate for a nominal active volume with \(\bar{N}_{\mathrm{In}}=5\). The entries quantify only the number of intentional In acceptors in the electrostatically active volume; the usable five-qubit-device yield will be lower because spacing, depth, gate lever arms, sensor visibility, and coupling hierarchy must also be favorable.}
\label{tab:poisson_occupancy}
\begin{tabular}{c c}
\toprule
\textbf{Occupancy criterion} & \textbf{Probability for \(\bar{N}_{\mathrm{In}}=5\)} \\
\midrule
\(N=3\) & 0.14 \\
\(N=4\) & 0.18 \\
\(N=5\) & 0.18 \\
\(N=6\) & 0.15 \\
\(N=7\) & 0.10 \\
\(4\leq N\leq6\) & 0.50 \\
\(3\leq N\leq7\) & 0.74 \\
\bottomrule
\end{tabular}
\end{table}

In the statistical layout model, each candidate device instance is represented by a set of acceptor positions
\begin{equation}
\{\mathbf{R}_1,\mathbf{R}_2,\dots,\mathbf{R}_N\},
\end{equation}
where \(N\) is drawn from the active-volume distribution and usable five-site subsets are selected by gate tuning and charge-sensing spectroscopy. This formulation makes the statistical-yield character of the first-generation architecture explicit.

A second design-level result is the negligibly small residual bulk disorder inside an individual acceptor-bound-hole volume. For a representative localization volume set by several effective Bohr radii,
\begin{equation}
V_q \sim (10~\mathrm{nm})^3 = 10^{-18}~\mathrm{cm^3},
\end{equation}
the expected number of unintended background impurities sampled by one qubit is
\begin{equation}
\langle N_{\mathrm{bg}}^{(q)} \rangle = n_{\mathrm{bg}}V_q \sim 10^{-8} \ll 1.
\label{eq:Nbg_q}
\end{equation}
Thus, the statistical layout module immediately places the platform in a favorable regime: the five-qubit chain is set by the intentional In array, while the uncontrolled detector-grade background is effectively absent at the single-qubit scale.

\paragraph{Nominal design estimate.}
For the nominal parameter set, Step~1 yields an active-volume occupancy of approximately five intentional In acceptors [Eq.~\eqref{eq:Nbar_In_active}], a characteristic inter-acceptor spacing near \SI{170}{\nano\meter}, and a residual bulk impurity occupancy per qubit volume far below unity [Eq.~\eqref{eq:Nbg_q}]. This establishes that the material system is compatible with a sparse, resolvable, statistically selected five-qubit register rather than a disorder-dominated impurity ensemble.

\subsection{Step 2: Electrostatic modeling}

The gate stack is modeled by solving the electrostatic potential $\phi(\mathbf{r})$ inside the device,
\begin{equation}
\nabla \cdot \left[\epsilon_r(\mathbf{r}) \epsilon_0 \nabla \phi(\mathbf{r}) \right] = -\rho(\mathbf{r}),
\label{eq:poisson}
\end{equation}
with the local electric field given by
\begin{equation}
\mathbf{E}(\mathbf{r}) = -\nabla \phi(\mathbf{r}).
\label{eq:efield}
\end{equation}
The relevant outputs are the vertical field at each acceptor site, the lateral field nonuniformity across the chain, and the crosstalk produced by neighboring plunger and barrier gates.

To quantify addressability, it is useful to define a gate-response matrix
\begin{equation}
\alpha_{ij} \equiv \frac{\partial E_i}{\partial V_j},
\label{eq:alpha_matrix}
\end{equation}
where $E_i$ is the local vertical field at qubit $i$ and $V_j$ is the voltage applied to gate $j$. A corresponding crosstalk metric is
\begin{equation}
C_{ij} \equiv \left|\frac{\alpha_{ij}}{\alpha_{ii}}\right|,
\qquad i\neq j.
\label{eq:crosstalk_metric}
\end{equation}

\paragraph{Nominal design estimate.}
For the nominal gate pitch matched to the acceptor spacing, the electrostatic solution is designed to yield a diagonally dominant response matrix, with adjacent-gate crosstalk typically remaining at the
\begin{equation}
C_{i,i\pm1} \lesssim 0.15\text{--}0.20
\end{equation}
level and more distant gate influence substantially smaller,
\begin{equation}
C_{ij}\lesssim 0.05 \qquad (|i-j|\ge 2).
\end{equation}
In the same nominal solution, the local plunger gates provide a Stark-tuning window large enough that neighboring qubit frequencies can be separated by more than the expected nearest-neighbor coupling scale. Thus, Step~2 indicates that individual site tuning is feasible without requiring fully isolated electrostatic dots.

\subsection{Step 3: Single-acceptor quantum modeling}

The field-dependent In-acceptor manifold is described using a valence-band acceptor Hamiltonian of the form
\begin{equation}
H_{\mathrm{acc}}
=
H_{\mathrm{KL}}
+
V_C(\mathbf{r})
+
V_{\mathrm{cc}}
+
H_{\epsilon}
+
e\,\phi(\mathbf{r})
+
H_B,
\label{eq:Hacc_workflow}
\end{equation}
where $H_{\mathrm{KL}}$ is the Kohn--Luttinger Hamiltonian, $V_C$ is the Coulomb potential of the acceptor, $V_{\mathrm{cc}}$ is the central-cell correction, $H_{\epsilon}$ is the strain contribution, and $H_B$ is the Zeeman interaction \cite{AbadilloUriel2016,AbadilloUriel2017,Terrazos2021}.

After projecting into the lowest Kramers doublet, the effective qubit Hamiltonian for site $i$ is written as
\begin{equation}
H_{q,i}
=
\frac{\mu_B}{2}\,\mathbf{B}\!\cdot\!\mathbf{g}_i(\mathbf{E}_i,\epsilon_i)\!\cdot\!\bm{\sigma}
+
\frac{\hbar\Omega_i(t)}{2}\sigma_i^x
+
\frac{\delta\omega_i}{2}\sigma_i^z.
\label{eq:Hq_i_workflow}
\end{equation}
From this projected model, the workflow extracts the qubit frequency
\begin{equation}
\omega_{q,i} = \frac{E_{1,i}-E_{0,i}}{\hbar},
\end{equation}
the effective $g$ tensor, the Stark sensitivity,
\begin{equation}
\chi_i^{(E)} \equiv \frac{\partial \omega_{q,i}}{\partial E_i},
\end{equation}
and the local dipole moment,
\begin{equation}
\mathbf{p}_i = e\,\Delta \mathbf{r}_i.
\end{equation}

\paragraph{Nominal design estimate.}
For the nominal bias range, the single-acceptor solver returns a well-defined lowest Kramers doublet at each site together with an anisotropic effective $g$ tensor,
\begin{equation}
g_{\parallel} \neq g_{\perp},
\end{equation}
and a finite gate-induced displacement of the acceptor-bound hole,
\begin{equation}
\Delta r_i \sim 0.5\text{--}1.0~\mathrm{nm},
\end{equation}
corresponding to a representative dipole magnitude
\begin{equation}
p_i = e\Delta r_i \sim (0.8\text{--}1.6)\times10^{-28}~\mathrm{C\,m}.
\label{eq:dipole_range}
\end{equation}
The extracted Stark sensitivity is sufficiently large that local gate tuning can shift the qubit frequency by several times the expected pairwise coupling strength, providing a practical route to frequency allocation and selective addressing. These outputs are consistent with the broader Ge-hole literature, which shows strong electric tunability, anisotropic $g$ tensors, and rapid electrically driven control in hole-based Ge qubits \cite{Hendrickx2020,Hendrickx2021Nature,Wang2022Hole,Hendrickx2024Sweet}.

\subsection{Step 4: Coupling extraction}

Once the local qubit parameters are known, pairwise interactions are extracted. Three channels are considered in the baseline workflow, and a fourth is added in the extended PnC branch.

\paragraph{(i) Dipole--dipole coupling.}
For qubits $i$ and $j$, the direct electrostatic interaction is stated in Eq.~\eqref{eq:dipole_dipole}.

For roughly uniform spacing, the next-nearest-neighbor dipolar coupling is suppressed by the cubic distance law,
\begin{equation}
\frac{U_{i,i+2}^{\mathrm{dd}}}{U_{i,i+1}^{\mathrm{dd}}}
\approx
\left(\frac{1}{2}\right)^3
=
\frac{1}{8},
\label{eq:dipole_hierarchy}
\end{equation}
so the dipolar channel naturally favors nearest-neighbor interactions within the selected chain.

Using the representative dipole range in Eq.~\eqref{eq:dipole_range}, a nearest-neighbor spacing
\begin{equation}
r_{i,i+1}\approx d_{\mathrm{avg}}\approx \SI{170}{\nano\meter},
\end{equation}
and $\epsilon_{\mathrm{Ge}}\approx 16\,\epsilon_0$, the nearest-neighbor dipolar coupling is estimated to be
\begin{equation}
\frac{U_{i,i+1}^{\mathrm{dd}}}{h}
\sim 1.1\text{--}4.4~\mathrm{MHz},
\label{eq:Udd_MHz}
\end{equation}
while the next-nearest-neighbor coupling is correspondingly
\begin{equation}
\frac{U_{i,i+2}^{\mathrm{dd}}}{h}
\sim 0.14\text{--}0.55~\mathrm{MHz}.
\label{eq:Udd_NNN}
\end{equation}

\paragraph{(ii) Exchange coupling.}
If neighboring wavefunctions are brought into appreciable overlap by unusually close spacing or gate-induced hybridization, the exchange scale is approximated as described by Eq.~\eqref{eq:Jex}.
Exchange is potentially the strongest short-range two-qubit interaction for close selected pairs, but it is also the most sensitive to position disorder and local electrostatics. For the nominal mean-spacing architecture, exchange should therefore be treated as an optional gate-activated or yield-dependent channel rather than as the default coupling mechanism.

\paragraph{Nominal design estimate.}
For the nominal geometry, the exchange channel is strongly suppressed in the idle configuration because the mean acceptor spacing is much larger than \(a_B^*\). It becomes significant only for close selected pairs or when barrier gates are used to lower the effective inter-site barrier and increase hybridization. In other words, the nominal estimate places the device in a regime where the static chain is primarily dipole-coupled or phonon-coupled, while exchange remains an optional yield-dependent and gate-activated channel for faster two-qubit operations.

\paragraph{(iii) Continuum phonon-mediated coupling.}
For a shared phonon mode of frequency $\omega_{\mathrm{ph}}$, the chain is described by
\begin{equation}
H_{\mathrm{bus}}
=
\hbar \omega_{\mathrm{ph}} a^\dagger a
+
\sum_i \frac{\hbar \omega_{q,i}}{2}\sigma_i^z
+
\sum_i \hbar g_i (a+a^\dagger)\sigma_i^x,
\label{eq:Hbus_workflow}
\end{equation}
where $g_i$ is an angular-frequency coupling rate and the factor $\hbar$ converts it to an energy in the Hamiltonian. This convention is used throughout when values such as $g_i/2\pi$ are quoted in megahertz. The Hamiltonian yields an effective interaction rate in the dispersive regime,
\begin{equation}
J_{ij}^{\mathrm{ph}} \sim \frac{g_i g_j}{\Delta},
\qquad
\Delta = \omega_{q}-\omega_{\mathrm{ph}}.
\label{eq:Jph_workflow}
\end{equation}
This channel is attractive because it can extend the coupling range beyond nearest-neighbor dipolar or exchange interaction and is especially natural in Ge hole-based systems, where strain coupling is intrinsically strong \cite{AbadilloUriel2023StrainSOI,Mei2025QST}.

\paragraph{Nominal design estimate.}
Using literature-motivated Ge-hole spin--phonon coupling targets in the
\begin{equation}
g/2\pi \sim 1\text{--}6~\mathrm{MHz}
\end{equation}
range and a representative dispersive detuning
\begin{equation}
\Delta/2\pi \sim 50~\mathrm{MHz},
\end{equation}
the phonon-mediated interaction scale is
\begin{equation}
J_{ij}^{\mathrm{ph}}/2\pi \sim 0.02\text{--}0.7~\mathrm{MHz}.
\label{eq:Jph_range}
\end{equation}
This is weaker than the nominal nearest-neighbor dipolar interaction but strong enough to be relevant for nonlocal coupling or quantum-bus operation in future phononic-crystal extensions.

\paragraph{(iv) PnC cavity branch.}
In the extended workflow, the local acoustic environment is replaced by a phononic crystal cavity or waveguide that reshapes the phononic density of states. The relevant acoustic eigenproblem is
\begin{equation}
\nabla \cdot \left[ \mathbf{C}(\mathbf{r}) : \nabla_s \mathbf{u}_n(\mathbf{r}) \right]
=
\rho(\mathbf{r}) \omega_n^2 \mathbf{u}_n(\mathbf{r}),
\label{eq:pnc_eigenproblem}
\end{equation}
where $\mathbf{u}_n(\mathbf{r})$ is the elastic displacement field, $\mathbf{C}(\mathbf{r})$ is the elastic tensor, and $\nabla_s$ denotes the symmetrized gradient. From these modes one extracts the cavity strain profile and the local zero-point strain,
\begin{equation}
\epsilon_{\alpha,i}^{\mathrm{zpf}},
\end{equation}
which in turn gives the cavity-enhanced qubit--phonon coupling
\begin{equation}
g_i^{\mathrm{PnC}}
\sim
\frac{1}{2\hbar}
\sum_{\alpha}
\lambda_{\alpha}\,
\epsilon_{\alpha,i}^{\mathrm{zpf}}.
\label{eq:g_pnc}
\end{equation}

A second key output of the PnC branch is the suppression factor
\begin{equation}
S_{\mathrm{PnC}}(\omega_q)
\equiv
\frac{\rho_{\mathrm{ph}}^{\mathrm{PnC}}(\omega_q)}
{\rho_{\mathrm{ph}}^{\mathrm{bulk}}(\omega_q)},
\label{eq:S_pnc}
\end{equation}
which quantifies how effectively the engineered acoustic bandgap suppresses unwanted environmental phonons at the qubit frequency.

\paragraph{Nominal design estimate.}
For a properly aligned PnC bandgap, the design target is
\begin{equation}
S_{\mathrm{PnC}}(\omega_q)\ll 1,
\end{equation}
with useful design-level suppression in the
\begin{equation}
S_{\mathrm{PnC}}\sim 10^{-2}\text{--}10^{-3}
\end{equation}
range. In the same extended branch, the cavity mode is chosen to preserve a selected localized strain field, maintaining
\begin{equation}
g_i^{\mathrm{PnC}}/2\pi \sim 1\text{--}6~\mathrm{MHz},
\end{equation}
while suppressing the unwanted acoustic continuum. Architecturally, this means that the PnC branch can improve phonon-limited relaxation and simultaneously sharpen the selectivity of phonon-mediated coupling. It is therefore best interpreted as a second-stage performance-enhancement path rather than a requirement for the baseline five-qubit chain.

\subsection{Step 5: Chain-level dynamics and disorder-ensemble analysis}

The final stage simulates the full five-qubit chain using an effective Hamiltonian of the form
\begin{equation}
H_{\mathrm{chain}}
=
\sum_{i=1}^{5}
\left[
\frac{\hbar \omega_{q,i}}{2}\sigma_i^z
+
\frac{\hbar \Omega_i(t)}{2}\sigma_i^x
\right]
+
\sum_{i<j}
\left(
J_{ij}^{xx}\sigma_i^x\sigma_j^x
+
J_{ij}^{zz}\sigma_i^z\sigma_j^z
\right),
\label{eq:Hchain_workflow}
\end{equation}
with the dynamical evolution optionally treated in open-system form,
\begin{equation}
\dot{\rho}
=
-\frac{i}{\hbar}[H_{\mathrm{chain}},\rho]
+
\sum_i
\left[
\frac{1}{T_{1,i}}\mathcal{D}[\sigma_i^-]\rho
+
\frac{1}{T_{\phi,i}}\mathcal{D}[\sigma_i^z]\rho
\right],
\label{eq:lindblad_workflow}
\end{equation}
where $\mathcal{D}[L]\rho = L\rho L^\dagger - \tfrac{1}{2}\{L^\dagger L,\rho\}$.

In the PnC-enhanced branch, the same master-equation framework is retained, but the phonon-limited relaxation channel is modified by the engineered local acoustic density of states. At the level of rates, this may be summarized schematically as
\begin{equation}
\Gamma_{1}^{\mathrm{ph,PnC}}
\approx
S_{\mathrm{PnC}}(\omega_q)\,
\Gamma_{1}^{\mathrm{ph,bulk}},
\label{eq:Gamma_PnC}
\end{equation}
provided the qubit frequency lies inside the phononic stop band and away from an intentionally coupled cavity resonance.

To assess robustness, the chain-level model is evaluated over an ensemble of disorder realizations. For each realization, the following metrics are extracted:
\begin{enumerate}
    \item frequency spread across the five qubits,
    \item achievable Stark tuning range,
    \item nearest-neighbor versus next-nearest-neighbor coupling ratio,
    \item crosstalk under single-qubit addressing,
    \item existence of a practical two-qubit operating window,
    \item and, in the extended branch, the degree of phonon-limited relaxation suppression and cavity-mode selectivity.
\end{enumerate}
The resulting ensemble distribution defines the \emph{design yield},
\begin{equation}
Y = \frac{N_{\mathrm{pass}}}{N_{\mathrm{tot}}},
\label{eq:yield}
\end{equation}
that is, the fraction of nominal device realizations satisfying the chosen operating criteria.

A useful addressability metric is
\begin{equation}
\mathcal{A}_{ij} \equiv \frac{|\omega_{q,i}-\omega_{q,j}|}{J_{ij}},
\label{eq:addressability_metric}
\end{equation}
for which idle operation requires
\begin{equation}
\mathcal{A}_{ij} \gg 1
\end{equation}
for spectator pairs, while selective two-qubit operation is approached by tuning one chosen pair toward
\begin{equation}
\mathcal{A}_{ij}\sim 1.
\end{equation}

\paragraph{Nominal design estimate.}
For the nominal parameter set, the disorder-ensemble analysis places the chain in a favorable control hierarchy:
\begin{enumerate}
    \item spectator qubits remain frequency-resolved in the idle configuration,
    \item nearest-neighbor interactions dominate over next-nearest-neighbor interactions,
    \item at least one pairwise entangling window can be opened by barrier and plunger retuning without collapsing the entire chain into a globally resonant state.
\end{enumerate}
In design terms, the chain therefore operates in the desired regime
\begin{equation}
J_{i,i+2} \ll J_{i,i+1} \ll |\omega_{q,i}-\omega_{q,j}|_{\mathrm{idle}},
\label{eq:control_hierarchy}
\end{equation}
with the middle inequality intentionally relaxed for a selected pair during gate-enabled two-qubit operation.

In the PnC-enhanced branch, the same control hierarchy is preserved while the phonon-limited relaxation floor is reduced and the phonon-mediated channel becomes more spectrally selective. Thus, the extended workflow does not alter the basic control logic of the five-qubit device; rather, it improves the coherence and coupling environment around that same logic.

\subsection{Step-by-step nominal design estimates}

The most important outputs of the workflow are summarized directly by modeling stage in Table~\ref{tab:workflow_results}.

\begin{table}[H]
\centering
\caption{Step-by-step nominal design estimates for the five-qubit In-acceptor platform in detector-grade Ge, including the engineered PnC-enhanced branch. These values are intended as self-consistent design-study outputs rather than final experimentally validated benchmarks.}
\label{tab:workflow_results}
\begin{tabular}{p{3.0cm}p{4.0cm}p{5.5cm}}
\toprule
\textbf{Modeling step} & \textbf{Representative output} & \textbf{Interpretation} \\
\midrule
Materials / layout
& $d_{\mathrm{avg}}\approx \SI{170}{nm}$, $\bar{N}_{\mathrm{In}}\sim 4.5$--$5.8$, $\langle N_{\mathrm{bg}}^{(q)}\rangle\sim 10^{-5}$
& Statistically selected five-site register in the defined active volume; bulk background effectively absent at single-qubit scale \\

Electrostatics
& $C_{i,i\pm1}\lesssim 0.15$--$0.20$, $C_{ij}\lesssim 0.05$ for $|i-j|\ge2$
& Gate response is diagonally dominant; local Stark tuning remains feasible \\

Single-acceptor solver
& $\Delta r_i\sim 0.5$--$1.0~\mathrm{nm}$, $p_i\sim (0.8$--$1.6)\times10^{-28}~\mathrm{C\,m}$, anisotropic $g$ tensor
& Finite dipole moment and strong electric tunability enable all-electrical control \\

Coupling extraction
& $U_{i,i+1}^{\mathrm{dd}}/h\sim 1.1$--$4.4~\mathrm{MHz}$, $U_{i,i+2}^{\mathrm{dd}}/h\sim 0.14$--$0.55~\mathrm{MHz}$; phonon-bus target $J_{ij}^{\mathrm{ph}}/2\pi\sim 0.02$--$0.7~\mathrm{MHz}$
& Nearest-neighbor interactions dominate; phonon coupling remains a useful extension path \\

PnC cavity branch
& $S_{\mathrm{PnC}}\sim 10^{-2}$--$10^{-3}$, selected $g_i^{\mathrm{PnC}}/2\pi\sim 1$--$6~\mathrm{MHz}$
& Unwanted phonon continuum suppressed while selected cavity-mediated coupling is preserved \\

Chain dynamics / yield
& Idle hierarchy $J_{i,i+2}\ll J_{i,i+1}\ll |\omega_{q,i}-\omega_{q,j}|_{\mathrm{idle}}$; selective pair tuning possible
& The chain is addressable in idle mode and tunable into pairwise entangling windows \\
\bottomrule
\end{tabular}
\end{table}

\subsection{Design-level conclusions from the modeling workflow}

At the design level, the improved modeling workflow yields six concrete conclusions.

First, the target indium concentration and active-volume definition generate candidate active volumes with an average occupancy near five acceptors over a \SI{1}{\micro\meter} channel, without requiring deterministic single-ion implantation. Usable five-site registers are then identified by post-fabrication mapping and electrical selection.

Second, the detector-grade background is sufficiently low that uncontrolled bulk disorder is negligible at the qubit scale, making the local gate environment rather than the host crystal the dominant source of device-to-device variation.

Third, the electrostatic design remains in a diagonally dominant tuning regime, so individual plunger gates can be used for local qubit tuning without destroying addressability.

Fourth, the extracted coupling hierarchy is favorable for a one-dimensional processor tile: nearest-neighbor dipolar coupling lies naturally in the MHz regime, next-nearest-neighbor dipolar terms are suppressed by geometry, exchange can be activated selectively, and phonon-mediated coupling remains available as a longer-range extension path.

Fifth, the engineered PnC branch shows that phononic crystal cavity engineering can, in principle, improve performance without changing the basic device logic. Its main benefits are suppression of unwanted environmental phonons and increased selectivity of the cavity-mediated coupling channel.

Sixth, the chain-level dynamics support the basic operating logic of the architecture: a frequency-resolved idle state, local electrical control, and gate-enabled pairwise interaction windows. In that sense, the nominal five-qubit In-acceptor chain represents a physically motivated mesoscopic device concept rather than a purely illustrative schematic: its materials, electrostatic, single-qubit, and coupling scales are mutually compatible, with the PnC cavity providing a clear pathway to further performance enhancement.

\section{Fabrication Strategy}
\label{sec:fabrication_strategy}

The proposed five-qubit In-acceptor platform is intended to be fabricated through a staged process that combines detector-grade Ge crystal growth, controlled indium incorporation, low-damage surface preparation, lithographically defined gate/readout integration, and---in an engineered second-stage branch---phononic crystal cavity (PnC) patterning. The central fabrication philosophy is to separate the problem into three layers: first, create an ultra-low-disorder detector-grade Ge host with a calibrated intentional acceptor density; second, add only the minimum gate and sensing structures needed for tuning, coupling control, and readout; and third, if enhanced phonon engineering is desired, pattern the local acoustic environment so that unwanted phonons are suppressed while selected cavity or waveguide modes are preserved. This is different from a conventional gate-defined quantum-dot platform, in which the full confinement potential must be created electrostatically. Here, the acceptor sites define the qubits, the gate stack primarily provides tuning and measurement functionality, and the PnC---when added---acts as an acoustic filter and cavity interface rather than as the primary qubit-defining structure.

\subsection{Stage I: detector-grade Ge purification and intentional indium incorporation}

The starting material is detector-grade high-purity germanium (HPGe), prepared by zone refining followed by Czochralski growth in a hydrogen atmosphere. This route is already well established for large HPGe crystal production and has been demonstrated to yield material with controlled dislocation density suitable for detector fabrication and advanced device development \cite{Wang2012HPGe,Wang2014Dislocation}. For the present platform, the initial target is the residual background impurity concentration given in Eq.~\eqref{eq:nbg}, so that uncontrolled bulk electrostatic and strain disorder remain negligible compared with the intentional acceptor array.

Intentional In incorporation is then introduced during crystal growth. A key practical advantage is that the electrical properties and distribution of indium in Ge have already been measured experimentally, including an effective segregation coefficient
\begin{equation}
k_{\mathrm{eff}}^{\mathrm{(In)}} \simeq 9\times 10^{-4},
\label{eq:keff_in_fab}
\end{equation}
over a broad concentration range \cite{Wang2018In}. To first approximation, the axial solid concentration follows a Scheil-type form
\begin{equation}
C_s(f) \approx k_{\mathrm{eff}} C_0 (1-f)^{k_{\mathrm{eff}}-1},
\label{eq:scheil}
\end{equation}
where $C_0$ is the initial melt concentration and $f$ is the solidified fraction. Because $k_{\mathrm{eff}}$ is very small, the In concentration is not expected to be uniform over an entire boule. Instead, the fabrication strategy is to use the calibrated incorporation behavior to identify and cut wafers from a narrow axial region where the local indium concentration lies near the target value in Eq.~\eqref{eq:nin}. Thus, the relevant fabrication objective is not global uniformity over the full crystal, but local compositional control over the wafer area used for the qubit device. Growth variables such as initial melt concentration, solidified fraction, pulling conditions, rotation, and thermal-gradient/interface control can be used to tune the average concentration profile, but they do not determine the exact lateral positions or depths of individual In acceptors.

\subsection{Stage II: wafer selection, orientation, and materials qualification}

After growth, wafers should be selected from crystal regions that simultaneously satisfy three criteria: low dislocation density, the desired local In density, and low radial/axial compositional variation across the device area. In practice, this requires a combination of Hall-effect measurements, resistivity mapping, and defect/dislocation screening, together with standard crystal slicing, lapping, and chemo-mechanical polishing \cite{Wang2014Dislocation,Wang2018In}. The relevant outcome of this stage is not just a polished Ge substrate, but a pre-qualified qubit wafer whose local material properties are already known before nanofabrication begins.

A useful intermediate step is to fabricate simple planar or Hall-bar-style test structures on companion wafers from the same growth region. Detector-style planar structures are especially valuable here because they provide a direct probe of crystal quality, leakage behavior, and contact performance. In particular, planar detectors have already been fabricated on USD-grown HPGe crystals by coating the Ge surface with a high-resistivity amorphous-Ge (a-Ge) layer followed by a thin Al layer to define the contact area \cite{Wei2018aGeContacts}. Although such detector contacts are not themselves the final qubit-gate architecture, they provide an experimentally grounded route for early wafer qualification and process calibration before the more delicate qubit devices are fabricated.

If a phononic-crystal branch is to be pursued later, this qualification stage should also identify wafers and device regions suitable for local membrane thinning or patterned acoustic structuring. In other words, the same wafer-screening process used to qualify the qubit material should also screen for low-defect regions likely to tolerate the added nanofabrication burden of a PnC cavity.

\subsection{Stage III: surface preparation, gate-stack formation, and engineered PnC patterning}

Once a wafer has been qualified, the baseline qubit device is fabricated on a freshly prepared low-damage surface. The essential requirement at this stage is to preserve the low-disorder bulk advantage of the detector-grade Ge while adding a lithographically defined gate stack that can tune local electric fields without introducing excessive interface charge noise. In contrast to a quantum-dot processor, the gates do not need to create the qubits; instead, they tune the acceptor-bound-hole spectrum, Stark shift, and inter-qubit coupling. This substantially relaxes the electrostatic design burden.

The proposed baseline gate architecture consists of local plunger gates above each nominal qubit site and barrier gates between neighboring sites. In the simplest implementation, the plunger gates control the local vertical electric field and therefore tune the qubit frequency through Stark shifts and electric-dipole spin resonance (EDSR)-active admixture, while the barrier gates adjust the electrostatic environment between neighboring acceptors. This design philosophy is closely aligned with planar Ge hole-spin devices, where integrated charge sensing, RF reflectometry, and all-electrical control have already been demonstrated \cite{Hendrickx2020,Hendrickx2021Nature,Scappucci2021}. For the present acceptor platform, however, the electrostatic task is simpler because the qubit is defined by the impurity site rather than by a gate-created dot.

An engineered second branch of this stage is the fabrication of a phononic crystal cavity. The role of the PnC is not to define the qubits, but to reshape the acoustic environment seen by them: unwanted environmental phonons can be suppressed by an engineered acoustic bandgap, while selected cavity or waveguide modes can be preserved for controlled spin--phonon coupling \cite{Smelyanskiy2014GePhononic,Mei2025QST}. If this branch is pursued, the most natural route is to form a locally thinned Ge membrane or suspended acoustic region near the qubit chain and then pattern a periodic hole lattice or equivalent elastic structure using high-resolution lithography and anisotropic dry etching. A defect cavity or short phononic waveguide may then be introduced as a controlled break in the periodic pattern.

From a fabrication standpoint, the PnC branch introduces two additional constraints. First, the patterned acoustic region must be aligned relative to the qubit chain so that the relevant cavity strain field overlaps the acceptor sites without placing the qubits too close to rough etched boundaries. Second, the etch and release processes must preserve the low-disorder electrical environment established in the preceding stages. For this reason, the PnC should be treated as a deliberate second-stage enhancement of the platform rather than as part of the minimum viable first-generation device.

\subsection{Stage IV: readout integration, interconnects, and PnC-compatible packaging}

The initial five-qubit devices should incorporate a nearby charge-sensitive structure for spin-to-charge readout. A practical route is to adopt readout concepts already proven in planar Ge quantum devices, including separate charge sensors and radio-frequency reflectometry \cite{Hendrickx2020}. The sensor can be placed adjacent to one end of the five-qubit chain or coupled capacitively to an intermediate node, depending on the final circuit layout. For early prototypes, the main goal is not yet full parallel readout of all five qubits, but reliable demonstration of local state discrimination, field tuning, and pairwise coupling control.

For the baseline five-qubit device, the preferred first-generation readout pathway is spin-to-charge conversion followed by radio-frequency charge sensing. A nearby RF-QPC, RF-SET, or gate-based dispersive charge sensor can be capacitively coupled to one end of the acceptor chain or to an intermediate sensing node. The sensor does not need to define the qubit; its role is to detect whether a selected acceptor has changed charge configuration after a spin-dependent mapping pulse. This separation between qubit definition and charge sensing is important because the acceptor provides the localized bound-hole state, while the lithographic sensor provides only state discrimination.

Two complementary spin-to-charge conversion mechanisms are possible. In the simplest single-acceptor readout mode, a gate pulse maps the spin state onto a spin-dependent tunneling or ionization event between the acceptor and a nearby reservoir or charge-sensing island. The resulting charge transition is detected by RF reflectometry. In a two-acceptor mode, neighboring acceptors can be pulsed into a regime where Pauli selection rules or spin-dependent hybridization convert the spin configuration into a measurable charge occupation difference, analogous to Pauli-spin-blockade readout in gate-defined spin qubits. The first-generation goal is not simultaneous high-fidelity readout of all five qubits, but demonstration that at least one selected acceptor, and subsequently one selected pair, can be initialized, tuned, manipulated, and read out with sufficient signal-to-noise.

This readout strategy also provides a practical method for mapping the statistically defined acceptor chain. By sweeping local plunger and barrier gates while monitoring the RF charge sensor, one can identify charge transitions, estimate lever arms, assign acceptor sites to local gates, and determine which subset of acceptors forms a usable five-site register. Such measurements provide an effective gate-response map rather than a direct three-dimensional atomic map: lateral position can be constrained from the pattern of relative gate couplings, while depth below the gate is generally inferred only with support from electrostatic simulations or calibration structures. Thus, readout is not only a measurement tool but also part of the device-screening and yield-selection workflow for the proposed statistical In-acceptor architecture.

At the packaging level, the baseline device should be mounted in a geometry compatible with high-frequency gate excitation, low-noise DC biasing, and cryogenic charge-sensing measurements. Because the proposed architecture is electrically controlled, the most critical interconnect requirement is low-crosstalk routing of the plunger and barrier gates rather than the integration of large on-chip magnetic structures. This is again a practical advantage of an acceptor-based Ge platform.

If the PnC branch is added, packaging must also be made compatible with the mechanical quality factor of the cavity or membrane. In that case, the chip mount, wire-bond geometry, and thermal anchoring strategy should be co-designed to minimize clamping loss and unintentional acoustic loading of the patterned region. Thus, PnC-compatible packaging is not a wholly separate challenge, but an extension of the same RF and cryogenic design discipline already required for the baseline device.

\subsection{Stage V: staged experimental validation}

A realistic experimental program should proceed in three baseline generations, followed by an engineered PnC-enhanced branch. The first generation should focus on materials qualification structures, including Hall devices, planar detector-like contacts, and simple capacitive test structures, in order to verify the local In concentration, leakage behavior, and interface quality. The second generation should target one- and two-qubit devices to establish Stark tuning, single-qubit control, and nearest-neighbor coupling windows. The third generation should implement the full five-qubit chain with integrated sensing and local gate control.

Only after this baseline chain has been established should the PnC-enhanced branch be introduced. In that branch, the same few-qubit or five-qubit architecture is reimplemented within a patterned phononic environment in order to test three specific gains: suppression of unwanted phonon-limited relaxation, retention of a selected cavity or waveguide mode, and enhancement of phonon-mediated coupling selectivity. This staged sequence is important because it preserves the experimental clarity of the program: the baseline acceptor platform is validated first, and the PnC is then introduced as a performance-enhancing architectural layer rather than as an uncontrolled additional complication.

This staged strategy is important because the proposed platform combines three technological lineages that are individually mature but not yet fully integrated: detector-grade Ge crystal growth, planar Ge hole-qubit device control, and phononic-cavity engineering. The fabrication challenge is therefore primarily one of integration rather than of inventing an entirely new materials platform from scratch. In that sense, the fabrication path is experimentally credible: the crystal-growth basis, the In-incorporation data, the detector-contact processing, the gate-controlled Ge spin-qubit methodology, and the phononic-cavity design logic all already exist in adjacent forms \cite{Wang2012HPGe,Wang2018In,Wei2018aGeContacts,Hendrickx2020,Hendrickx2021Nature,Smelyanskiy2014GePhononic,Mei2025QST}. The staged fabrication and validation sequence is summarized in Table~\ref{tab:fabrication_strategy}.

\begin{table}[H]
\centering
\caption{Proposed staged fabrication strategy for the five-qubit In-acceptor platform in detector-grade Ge, including the engineered phononic-crystal cavity branch.}
\label{tab:fabrication_strategy}
\begin{tabular}{p{3.2cm}p{4.8cm}p{5.6cm}}
\hline
\textbf{Stage} & \textbf{Primary objective} & \textbf{Representative outputs} \\
\hline
Crystal purification and growth
& Produce detector-grade Ge with controlled dislocation density and calibrated In incorporation
& HPGe boule, local $n_{\mathrm{bg}}$, local $n_{\mathrm{In}}$, low-dislocation growth region \\

Wafer selection and qualification
& Select device-grade wafers with acceptable composition and defect density
& Hall data, resistivity map, dislocation screening, polished qubit wafer \\

Surface preparation and baseline gate-stack fabrication
& Form a low-damage surface and define plunger/barrier gate architecture
& Patterned gate stack for local Stark tuning and coupling control \\

engineered PnC patterning
& Add acoustic bandgap structure and, if desired, a defect cavity or waveguide
& Patterned phononic region, selected cavity mode, filtered acoustic environment \\

Readout integration and packaging
& Add charge-sensitive measurement structure and cryogenic interconnects
& RF-compatible sensor, low-noise gate routing, packaged device, PnC-compatible mount if needed \\

Staged validation
& Progress from materials test structures to baseline few-qubit devices, then to PnC-enhanced prototypes
& Qualified process flow, single-/two-qubit benchmarks, five-qubit prototype, engineered cavity-enhanced device \\
\hline
\end{tabular}
\end{table}

\section{Scalability to Longer One-Dimensional Buses and Larger Arrays}
\label{sec:scalability}

The five-qubit chain proposed here is not intended as an endpoint, but as the minimal experimentally meaningful building block of a broader Ge acceptor-spin architecture. Its real value lies in establishing the physical regime needed for scaling: atom-defined qubit localization in an ultra-low-disorder host, local electric-field control through a compact gate stack, nearest-neighbor coupling that is strong enough for entangling operations yet sufficiently local for addressability, and a materials platform compatible with future phononic or photonic interconnects.

Two features of the present platform are particularly favorable for scaling. First, detector-grade Ge suppresses the uncontrolled bulk electrostatic and strain background to a level where local gate fields, rather than random host disorder, should dominate the qubit Hamiltonian. Second, the qubit is defined by the acceptor site rather than by a fully gate-created quantum dot. This reduces the electrostatic burden of scaling: larger structures do not require constructing every confinement potential from scratch, but rather tuning and coupling a pre-existing impurity-defined qubit network.

\subsection{Scaling from a five-qubit chain to a one-dimensional quantum bus}

At the design level, the extension from five qubits to a longer one-dimensional register is conceptually straightforward. If the target In concentration remains near the value in Eq.~\eqref{eq:nin}, then the mean spacing remains near the value derived in Eq.~\eqref{eq:davg}. The corresponding number of qubit sites supported by a one-dimensional active region of length $L$ is approximately
\begin{equation}
N_{1\mathrm{D}} \sim \frac{L}{d_{\mathrm{avg}}}.
\label{eq:N1D_scaling}
\end{equation}
Thus,
\begin{equation}
L=\SI{1}{\micro\meter} \rightarrow N_{1\mathrm{D}}\sim 5,
\qquad
L=\SI{5}{\micro\meter} \rightarrow N_{1\mathrm{D}}\sim 29,
\qquad
L=\SI{10}{\micro\meter} \rightarrow N_{1\mathrm{D}}\sim 59.
\end{equation}
Even allowing for practical spacing fluctuations and layout margins, these estimates show that detector-grade Ge with locally calibrated intentional In incorporation could in principle support medium-length one-dimensional qubit buses on the \SI{5}{\micro\meter}--\SI{10}{\micro\meter} scale.

The principal challenge in such a bus is not simply placing more acceptors, but preserving a useful interaction hierarchy. For a dipolar interaction,
\begin{equation}
U_{ij}^{\mathrm{dd}} \propto \frac{1}{r_{ij}^{3}},
\end{equation}
so nearest-neighbor coupling remains dominant provided the spacing disorder is moderate. In an approximately uniform chain,
\begin{equation}
\frac{U_{i,i+m}^{\mathrm{dd}}}{U_{i,i+1}^{\mathrm{dd}}}\approx \frac{1}{m^3},
\end{equation}
which naturally suppresses long-range spectral crowding. A useful figure of merit for one-dimensional scaling is
\begin{equation}
\Lambda_{1\mathrm{D}} \equiv \frac{J_{i,i+1}}{\delta\omega_{\mathrm{idle}}},
\end{equation}
where $\delta\omega_{\mathrm{idle}}$ denotes the residual site-to-site qubit-frequency spread after static tuning. For scalable operation, one typically wants $\Lambda_{1\mathrm{D}}<1$ in the idle configuration so that qubits remain frequency-resolved, while local retuning can bring a selected pair toward resonance for two-qubit operations.

\subsection{Role of phonon-mediated coupling and phononic crystal cavities}

Although nearest-neighbor dipolar and exchange interactions are sufficient for short chains, longer one-dimensional buses benefit from an interaction channel that can bypass purely local routing. Because acceptor-bound holes in Ge are intrinsically strain-active and spin--orbit-active, they are naturally suited to phonon-mediated coupling and to hybrid phononic architectures. A generic bus Hamiltonian may be written as
\begin{equation}
H_{\mathrm{bus}}
=
\hbar \omega_{\mathrm{ph}} a^\dagger a
+
\sum_i \frac{\hbar \omega_{q,i}}{2}\sigma_i^z
+
\sum_i \hbar g_i (a+a^\dagger)\sigma_i^x,
\end{equation}
where $g_i$ is again an angular-frequency coupling rate. The corresponding dispersive interaction rate is
\begin{equation}
J_{ij}^{\mathrm{ph}} \sim \frac{g_i g_j}{\Delta},
\qquad
\Delta=\omega_q-\omega_{\mathrm{ph}}.
\end{equation}
The key systems-level point is that $J_{ij}^{\mathrm{ph}}$ need not decay as rapidly with physical separation as direct electrostatic coupling.

At this longer-distance scale, a phononic crystal cavity (PnC) or guided phononic waveguide becomes especially useful. In the baseline five-qubit chain, a PnC is not strictly required because local electrical tuning and short-range coupling already define a meaningful device. In a longer bus, however, the acoustic environment increasingly becomes part of the scalability problem itself. A PnC addresses this in two complementary ways. First, it suppresses unwanted environmental phonons by opening an acoustic bandgap around the qubit frequency. Second, by introducing a defect cavity or engineered waveguide inside the phononic crystal, it preserves one or a few selected acoustic modes while suppressing the surrounding continuum \cite{Smelyanskiy2014GePhononic,Mei2025QST}. For the present In-acceptor platform, the most realistic interpretation is therefore not that a phonon bus is required from the beginning, but that the same material system is naturally compatible with such an extension once the shorter chain has been demonstrated.

\subsection{From one-dimensional buses to modular arrays}

A full large-scale processor is unlikely to be realized as one indefinitely long acceptor chain. As in many semiconductor qubit architectures, the more realistic strategy is modular scaling: use compact, locally controllable one-dimensional units as building blocks and connect them through shared sensors, phononic links, capacitive couplers, or cryogenic routing layers. In this picture, the five-qubit chain functions as a primitive module, a longer \SI{5}{\micro\meter}--\SI{10}{\micro\meter} chain functions as a mesoscopic quantum bus, and larger systems are assembled from these buses rather than grown monolithically from a single tuning problem.

A useful scaling relation for a modular array is
\begin{equation}
N_{\mathrm{tot}} = N_{\mathrm{mod}} \times N_{\mathrm{bus}},
\end{equation}
where $N_{\mathrm{mod}}$ is the number of modules and $N_{\mathrm{bus}}$ is the number of qubits per one-dimensional bus. If a \SI{5}{\micro\meter} bus reliably supports $\sim 25$--30 sites, then even a modest four-module architecture would already enter the $\mathcal{O}(10^2)$-qubit regime. At that point the dominant challenges are no longer microscopic acceptor physics alone, but control wiring, calibration overhead, readout multiplexing, and module-to-module coupling.

\subsection{Scalability bottlenecks and outlook}

Several bottlenecks must still be overcome before this scaling pathway becomes realistic. The intentional acceptor profile must remain sufficiently uniform over longer distances; interface disorder becomes progressively more important as the chain length increases; calibration overhead grows with system size; and the long-term roles of exchange, dipole, continuum phonons, and PnC-enhanced cavity modes must be co-designed rather than treated independently. If a PnC cavity is added, fabrication complexity and packaging constraints also increase.

Taken together, the scaling picture is encouraging but still conditional. The present five-qubit architecture is already a meaningful unit because it demonstrates the coexistence of atom-like localization, electrical tunability, and a local interaction hierarchy within detector-grade Ge. Extending the same density and control philosophy to longer one-dimensional structures leads naturally to the concept of a Ge acceptor quantum bus, while extension beyond a single chain points toward modular arrays in which multiple buses are linked through hybrid electrostatic, phononic, or photonic channels. The most realistic conclusion is therefore neither that the In-acceptor platform is immediately ready to compete with the most mature Ge hole-dot processors, nor that it should be viewed only as an isolated impurity curiosity. Rather, its strongest scaling path likely lies in the intermediate regime between those extremes: longer one-dimensional buses and modular hybrid arrays built on the combination of detector-grade material quality, impurity-defined qubit localization, and strong hole-based electric- and strain-response.

\section{Comparison with Donor and Gate-Defined Ge Platforms}
\label{sec:comparison_platforms}

The proposed In-acceptor platform should be understood not as an isolated alternative to the rest of the Ge qubit ecosystem, but as a distinct intermediate modality positioned between donor-based and gate-defined approaches. Donor qubits in Ge offer atom-like confinement and hybrid electron--nuclear functionality, while gate-defined Ge hole qubits currently provide the strongest experimental demonstration of processor-oriented operation. The In-acceptor concept combines elements of both: it retains chemistry-defined localization, as in donor systems, while inheriting the spin-$3/2$ valence-band physics and all-electrical controllability characteristic of hole-based platforms \cite{Scappucci2021,AbadilloUriel2016,AbadilloUriel2017}.

A further comparison point, which becomes increasingly important at the architectural level, is the role of the acoustic environment. In all three Ge-based platforms considered here, phonons are both an opportunity and a liability: they can mediate useful interactions, but they also open relaxation and decoherence channels. For this reason, phononic crystal cavity (PnC) engineering provides an additional axis of comparison beyond the usual metrics of coherence, controllability, and lithographic complexity. Donor platforms already have a clear theoretical connection to Ge-based phononic crystals, while Ge hole-based platforms---including the present acceptor proposal---are especially attractive for PnC integration because of their strong strain and spin--orbit response \cite{Smelyanskiy2014GePhononic,Mei2025QST}.

\subsection{Comparison with Ge donor qubits}

The closest conceptual relative of the proposed In-acceptor chain is the Ge donor platform. In both cases, the qubit is hosted by an intentionally introduced substitutional impurity rather than a lithographically defined electrostatic dot. This gives both modalities an architectural advantage over purely gate-defined systems: the confinement potential is defined largely by chemistry, which can reduce one major source of lithographic variability. In donor-based Ge devices, this atom-like localization is combined with the possibility of hybrid electron--nuclear registers, strong electric tunability, and potentially more relaxed exchange-spacing requirements than in Si \cite{Sigillito2015,Sigillito2016,PicaLovett2016}.

The present In-acceptor platform differs from donor qubits in three fundamental ways. First, the qubit is hole-based rather than electron-based. Donor spins in Ge inherit the multivalley $L$-point conduction band and are therefore sensitive to valley-orbit structure, intervalley interference, and strong spin-lattice relaxation \cite{Sigillito2015,PicaLovett2016}. By contrast, acceptor qubits are built from the valley-free $\Gamma$-point valence manifold and avoid the conduction-band valley complications that affect donor and gate-defined electron devices \cite{Scappucci2021}. Second, the acceptor-bound hole is intrinsically spin--orbit-active, so the present platform supports all-electrical control through electric-dipole and strain-mediated channels that are much stronger than in donor-electron systems \cite{AbadilloUriel2016,AbadilloUriel2017}. Third, the strongest design burden shifts accordingly: donor qubits are limited mainly by spin-lattice relaxation and multivalley complexity, whereas acceptor qubits are more sensitive to local symmetry breaking, interface fields, and strain disorder.

This distinction becomes especially relevant when one includes PnC engineering. Donor qubits in Ge have an important conceptual advantage in this context because Ge-based PnC architectures were proposed early as a way to suppress unwanted phonon relaxation while preserving useful cavity-assisted coupling channels \cite{Smelyanskiy2014GePhononic}. However, the present In-acceptor concept may be even more naturally matched to a PnC-enhanced architecture because the qubit itself is built from a strain-active hole state. Donors remain attractive when atom-like reproducibility and hybrid memory functionality are the highest priorities; the proposed acceptor platform becomes more compelling when one instead values stronger intrinsic strain response, larger electric-dipole activity, and closer compatibility with cavity-enhanced spin--phonon coupling.

\subsection{Comparison with gate-defined Ge hole qubits}

The most important benchmark for the present work is gate-defined Ge hole qubits. Among all Ge-based spin-qubit modalities, gate-defined hole devices are currently the most experimentally mature. They have progressed from the first Ge hole-spin qubit demonstrations to fast two-qubit logic, ultrafast coherent control, sweet-spot operation with improved coherence, coherent spin shuttling, and multi-qubit control in larger arrays \cite{Watzinger2018,Hendrickx2020Fast,Hendrickx2020,Hendrickx2021Nature,Wang2022Hole,Hendrickx2024Sweet,vanRiggelen2024}.

Their principal strength is the combination of fast all-electrical control and a scalable planar geometry. Their main remaining challenges are not questions of feasibility, but of scaling: charge noise, gate-stack variability, tuning overhead, and the management of bias-dependent $g$-tensor anisotropy in larger arrays \cite{Scappucci2021,Terrazos2021,Wang2021Optimal,Sarkar2025Disorder,Hendrickx2024Sweet}. Relative to that platform, the proposed In-acceptor chain offers a different trade-off. Its main potential advantage is reduced electrostatic complexity: the qubits themselves are not created by the gates, but already exist as impurity-defined localized states. The gate stack is then used primarily for Stark tuning, addressability, coupling control, and readout. The cost of this simplification is reduced flexibility, since an acceptor qubit is tied to the impurity site and its microscopic environment.

The PnC comparison sharpens this distinction further. Gate-defined Ge hole qubits are also natural candidates for phononic-crystal integration because they are built from the same strongly strain-active valence-band manifold, and recent modeling has shown that high-purity Ge hole-spin devices can benefit substantially from phononic cavity protection and cavity-mediated coupling \cite{Mei2025QST}. Thus, a PnC is not unique to the acceptor proposal. The real difference is architectural. In a gate-defined hole device, the electrostatic confinement problem and the phononic-environment problem must both be solved simultaneously. In the present In-acceptor platform, the qubit localization is already chemistry-defined, so the PnC can be introduced mainly as a coherence and coupling layer rather than as part of the qubit-definition problem itself.

\subsection{Comparison with gate-defined Ge electron qubits}

Electron-spin dots remain conceptually attractive because of their simpler spin-$1/2$ encoding, but in Ge they inherit the same multivalley $L$-point conduction-band structure that complicates donor-electron devices \cite{Scappucci2021}. As a result, they are expected to be more sensitive to valley splitting, interface disorder, and strain than hole-based devices. For this reason, gate-defined Ge electron qubits remain less experimentally developed than the hole platform and are not yet the leading Ge route to scalable processor hardware.

Relative to gate-defined electron dots, the proposed In-acceptor platform again benefits from being built on the valence-band side of Ge: it avoids valley complications and naturally supports strong electric and strain coupling. The trade-off is that acceptor holes are more complex than simple spin-$1/2$ electrons and are correspondingly more sensitive to symmetry-breaking perturbations. From a PnC standpoint, this same distinction remains relevant: electron platforms may benefit from phononic filtering mainly as a relaxation-suppression tool, whereas hole-based platforms can more naturally use the same phononic structure for both protection and purposeful cavity-mediated coupling.

\subsection{Architectural interpretation}

Taken together, these comparisons place the proposed In-acceptor platform in a clear architectural niche. It is not yet positioned to displace gate-defined Ge hole qubits as the most mature processor-oriented Ge technology, nor is it the most natural route to long-lived hybrid memory registers, where donor-based concepts remain compelling. Instead, its strongest case is as a hybrid impurity-defined hole platform that combines atom-like localization, all-electrical control, low gate complexity in one-dimensional geometries, and compatibility with phonon-assisted coupling.

The inclusion of phononic crystal cavities makes this niche more precise. Donor qubits provide the clearest precedent for impurity-based PnC protection and phonon-mediated coupling in Ge. Gate-defined Ge hole qubits provide the clearest precedent for high-performance hole-spin control and, increasingly, for PnC-enhanced hole-spin architectures. The present In-acceptor proposal sits between these two: it couples the chemistry-defined localization of impurity qubits to the strong strain response of Ge holes, making it especially attractive for medium-scale one-dimensional buses, hybrid spin--phonon devices, and modular architectures in which impurity-defined qubits are coupled through electrical and phononic channels rather than through purely local gate-defined dots.

\section{Conclusion}
\label{sec:conclusion}

In this work, we proposed and analyzed a germanium-based acceptor-spin platform built on detector-grade high-purity Ge with intentionally incorporated indium acceptors. The central design idea is straightforward: use detector-grade Ge to suppress uncontrolled bulk disorder, and use a controlled In concentration near $2\times10^{14}~\mathrm{cm^{-3}}$ to create a statistically selectable ensemble of atom-like acceptor sites. At this density, a \SI{1}{\micro\meter}-long active channel with an appropriate transverse mode volume hosts approximately five acceptors on average, providing candidate active volumes from which usable five-site registers can be identified after post-fabrication mapping, without requiring deterministic single-ion implantation.

\begin{figure}[H]
    \centering
    \includegraphics[width=0.96\textwidth]{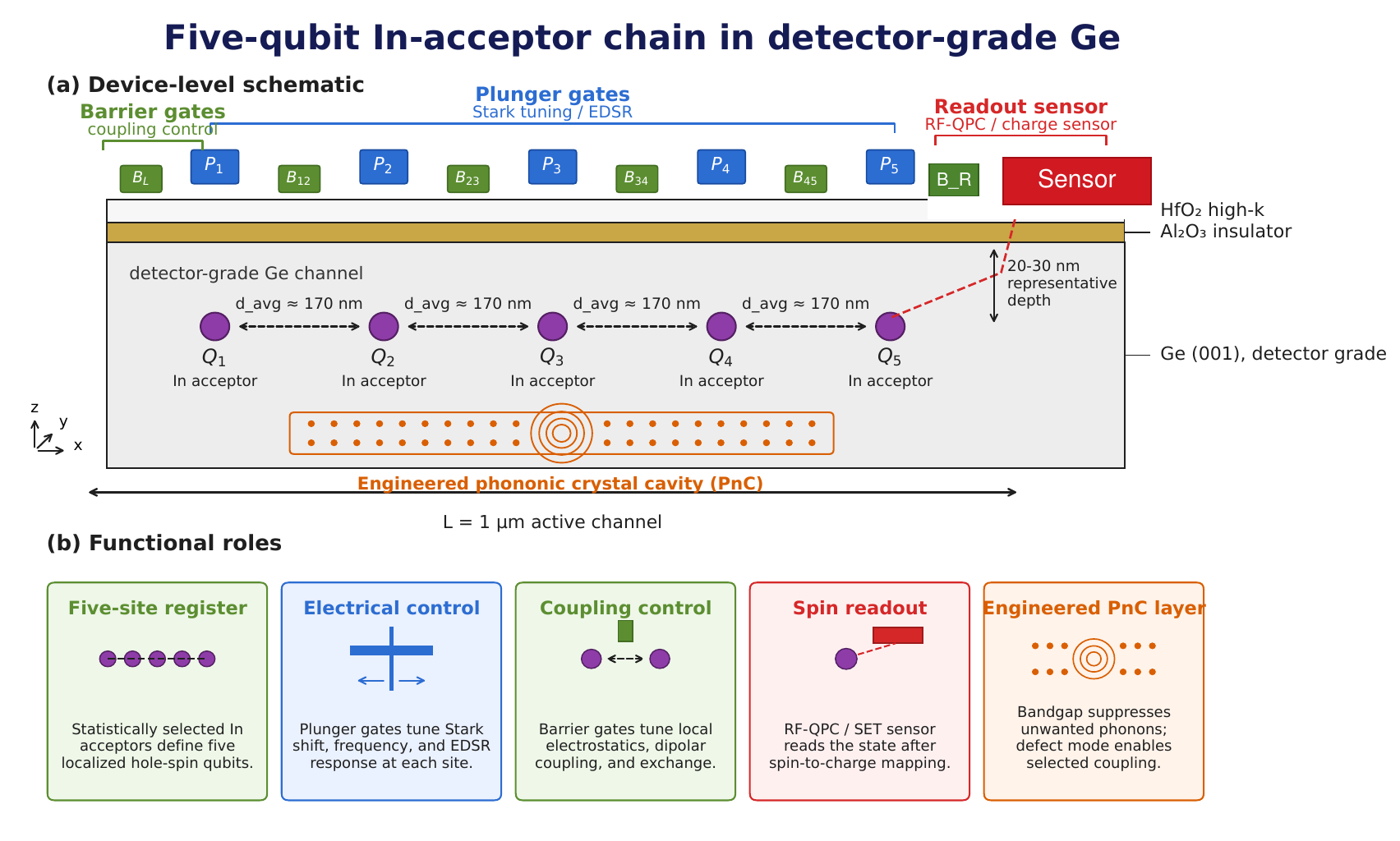}
    \caption{Conceptual summary of the proposed five-qubit In-acceptor platform in detector-grade Ge. \textbf{(a)} A \SI{1}{\micro\meter}-scale detector-grade Ge channel contains a statistically selected In-acceptor register, $Q_1$--$Q_5$, controlled by local plunger and barrier gates and read out through a nearby RF charge sensor. The labels denote one mapped post-fabrication realization, not predetermined implantation sites. \textbf{(b)} Functional modules of the architecture: statistical register selection, electrical control, coupling control, spin-to-charge readout, and optional PnC-enabled phonon filtering or selected cavity-mediated coupling.}
    \label{fig:five_qubit_pnc_conclusion}
\end{figure}

Figure~\ref{fig:five_qubit_pnc_conclusion} summarizes the architectural concept. A detector-grade Ge channel hosts a statistically selected five-qubit In-acceptor register with a characteristic three-dimensional acceptor spacing of approximately \SI{170}{\nano\meter}, while local plunger and barrier gates provide Stark tuning, coupling control, and access to spin-to-charge readout through a nearby RF-QPC or charge sensor. As in Fig.~\ref{fig:five_qubit_architecture}, any labels such as $Q_1$--$Q_5$ denote one successfully mapped post-fabrication realization rather than predetermined acceptor locations. The figure also captures the role of an engineered PnC: it does not define the qubits themselves, but can act as a second-stage enhancement by suppressing unwanted environmental phonons and enabling selected cavity-mediated spin--phonon coupling.

The proposed architecture occupies a distinct position within the Ge quantum landscape. Like donor platforms, it uses chemistry-defined localization; like gate-defined Ge hole qubits, it exploits spin--orbit-active valence-band physics for electrically tunable $g$ tensors, dipolar response, and all-electrical control \cite{Scappucci2021,AbadilloUriel2016,AbadilloUriel2017}. The result is a compact impurity-defined platform that could support mesoscopic one-dimensional registers, short quantum buses, and hybrid spin--phonon devices.

Our analysis indicates that the key materials requirements are compatible, at least at the design level, with existing HPGe growth capabilities. Residual impurity concentrations near $10^{10}~\mathrm{cm^{-3}}$ suppress the uncontrolled bulk strain and electrostatic background on the acceptor scale, while the intentional In concentration remains dilute enough for atom-like localization but high enough to produce a statistically selected five-site active volume. The resulting coupling hierarchy is deliberately conservative: plunger gates provide site-resolved Stark tuning, dipolar coupling provides the most robust mean-spacing interaction, exchange is reserved for close selected pairs or gate-enhanced hybridization, and phonon-mediated coupling remains a longer-range extension path.

The PnC branch should therefore be developed after, not before, the baseline chain. A first-generation device can test localization, tunability, addressability, spin-to-charge readout, and nearest-neighbor coupling without explicit phononic structuring. A later PnC layer can then reshape the acoustic density of states, reduce phonon-limited relaxation, and preserve selected cavity modes for controlled coupling \cite{Smelyanskiy2014GePhononic,Mei2025QST}.

The main open questions are the wafer-scale reproducibility of intentional In incorporation, the control of interface and gate-stack disorder, the statistical yield of usable five-site configurations, and the experimentally realized balance among dipolar, exchange, continuum-phonon, and cavity-mediated coupling channels. Resolving these questions will determine whether the platform can progress from a design concept to a competitive Ge quantum technology.

In summary, detector-grade Ge In-acceptor qubits offer a credible intermediate architecture between isolated impurity qubits and fully gate-defined Ge processors. Their appeal is the combination of impurity-defined localization, low-disorder materials, hole-based electrical control, and a clear staged pathway from baseline five-site devices to phonon-engineered quantum buses.

\section*{Acknowledgment}
This work was supported in part by NSF OISE 1743790, NSF PHYS 2117774, NSF OIA 2427805, NSF PHYS 2310027, NSF OIA 2437416, DOE DE-SC0024519, DE-SC0004768, and a research center supported by the State of South Dakota. 

\bibliographystyle{unsrtnat}
\bibliography{references}

\end{document}